\documentclass[printer]{aa}
\usepackage{graphicx,natbib,psfrag,amssymb,amsmath}
\usepackage{txfonts}
\usepackage{color}
\usepackage{textcomp}
\definecolor{noire}{rgb}{0,0,0} 
\definecolor{blue}{rgb}{0.1,0.5,0.7}
\bibpunct{(}{)}{;}{a}{}{,}
\usepackage{longtable,lscape} 
\usepackage{natbib}
\bibpunct{(}{)}{;}{a}{}{,}
\authorrunning{C. Fedeli \& A. Berciano Alba}
\titlerunning {Arcs produced by SMGs at radio and submm wavelengths}

\begin{document}

\title {On the abundance of gravitational arcs produced by
  submillimeter galaxies at radio and submm wavelengths}

\author{C. Fedeli\inst{1,2,3}\thanks{E-mail: cosimo.fedeli@unibo.it} \and A. Berciano Alba\inst{4,5}\thanks{E-mail: berciano@astro.rug.nl}}
 
 \institute{$^1$
     Dipartimento di Astronomia, Universit\`a di Bologna,
     Via Ranzani 1, I-40127 Bologna, Italy\\$^2$ INAF-Osservatorio
     Astronomico di Bologna, Via Ranzani 1, I-40127 Bologna, Italy\\$^3$
     INFN, Sezione di Bologna, Viale Berti Pichat 6/2, I-40127 Bologna, Italy\\$^4$
     Netherlands Institute for Radio Astronomy, Postbus 2, 7990 AA, Dwingeloo, The Netherlands\\$^5$
     Kapteyn Astronomical Institute, University of Groningen, PO Box 800, 9700 AV, Groningen,
     The Netherlands}
 
\date{\emph{Astronomy \& Astrophysics, submitted}}

\abstract{We predict the abundance of giant gravitational arcs
produced by submillimeter galaxies (SMGs) lensed by foreground galaxy
clusters, both at radio and submm wavelengths. The galaxy cluster
population is modeled in a realistic way with the use of semi-analytic
merger trees, while the density profiles of individual deflectors
take into account ellipticity and substructures. The adopted typical size of the radio and submm emitting
regions of SMGs is based on current radio/CO observations and the
FIR-radio correlation. The source redshift distribution has been
modeled using three different functions (based on spectroscopic/photometric
redshift measurements and a simple evolutionary model) to quantify the
effect of a high redshift tail on the number of arcs. The source
number counts are compatible with currently available observations,
and were suitably distorted to take into account the lensing
magnification bias. We present tables and plots for the numbers of
radio and submm arcs produced by SMGs as a function of
surface brightness, useful for the planning of future surveys
aimed at arc statistics studies. They show that e.g., the
detection of several hundred submm arcs on the whole sky with a
signal-to-noise ratio of at least $5$ requires a sensitivity of $1$ mJy
arcsec$^{-2}$ at $850~\mu$m. Approximately the same number of radio
arcs should be detected with the same signal-to-noise ratio with a
surface brightness threshold of $20~\mu$Jy arcsec$^{-2}$ at $1.4$
GHz. Comparisons of these results with previous work found in the
literature are also discussed.}

\maketitle

\section{Introduction}

The effect of gravitational lensing constitutes a unique research tool
in many astrophysical fields, since it allows one to investigate the
structures of both the lenses (e.g., galaxies and galaxy clusters) and
the lensed background sources, as well as to probe the three-dimensional mass distribution of the Universe. In particular, one of the most spectacular phenomena associated with
the gravitational deflection of light are the giant arcs observed in galaxy
clusters, which are caused by extended background sources lying in
the regions where the lensing magnification produced by the cluster is
strongest. The effective source magnification can easily reach $\sim 30$ in
these cases, providing the opportunity to detect and spatially resolve
the morphologies and internal dynamics of high redshift background
sources at a level of detail far greater than otherwise possible. An
illustration of this powerful technique was presented in
\cite{SW07.1}, where a magnification factor of $16$ by the cluster
RCS~0224-002 allowed them to study the star formation activity, mass and
feedback processes of a Lyman break galaxy at $z\sim5$, something that
(without the help of lensing) would not be possible beyond $z\sim2$
with current instruments.

A particularly interesting application of gravitational lensing is the
so-called arc statistics, i.e. the study of the abundance of large
tangential arcs in galaxy clusters. Among other things, this abundance
is sensitive to the cluster mass function, the cluster dynamical
activity (e.g., infall of matter, mergers, etc.) and internal
structure of host dark-matter halos (e.g., triaxiality and the
concentration of the density profile), which makes arc statistics a
unique tool to study the cluster population. Arc statistics
studies at optical and near-infrared wavelengths have been numerous in
the past decade on both the observational \citep{LU99.1,GL03.1,ZA03.1}
and theoretical sides \citep{BA94.1,WA98.1,BA03.2,ME03.1,WA05.1,FE08.1}.

All the currently known giant arcs come from detections in the
optical. However, the fraction of lensed sources observed in the
mm/submm wavebands is expected to be much larger than in the optical
\citep{BL96.1,BL97.1}: Due to the spectral shape of the thermal dust
emission, the observed submm flux density of dusty galaxies with a
given luminosity remains approximately constant in the redshift range
$1 \lesssim z \lesssim 8$ instead of declining with increasing
distance (usually referred to as "negative K-correction", see
\citealt{BL93.1,BL96.2,BL02.1}). This effect, together with the steep
slope of the observed submm number counts, produces a strong
magnification bias that makes submm galaxies (hereafter SMGs) an ideal
source population for the production of lensed arcs.

The SMGs were first detected about a decade ago with
SCUBA\footnote{Submillimeter Common User Bolometer Array
\citep{HO99.1}, which used to be mounted at the James Clerk Maxwell
Telescope (JCMT) located in Hawaii}
\citep{SM97.1,HU98.1,BA98.2,EA99.1}. The current observational
evidence indicates that these objects are high-redshift dust obscured
galaxies, in which the rest frame FIR peak of emission is observed in
the submm band. Their FIR luminosities, in the range $10^{11}-10^{13}
L_\odot$, are $\sim 100$ times higher than what is observed in local
spirals.  Their energy output seems to be dominated by star formation
processes induced by galaxy interactions/mergers, although a good
fraction ($\sim 30-50 \%$) of SMGs also seems to host an Active
Galactic Nucleus \citep[AGN, e.g.][]{AL05.2, MI09.1}. The
available evidence also suggests that SMGs might be the progenitors of
massive local ellipticals \citep{LI99.1, SM02.1, SM04.1, WE03.1,
GE03.1, AL03.1, AL05.2, SW06.1,SW08.1, MI09.1}.

In the $\sim 30$ clusters observed with SCUBA
\citep{SM02.1,CH02.2,CO02.1,KN08.2}, only 4 multiply imaged SMGs have
been reported to date \citep{BO04.1,KN04.1,KN08.2}. This extremely low
detection rate is due to three major limitations of current
$850~\mu$m surveys: (i) very small sky coverage ($\sim 3$ square
degrees, including all cluster and blank field surveys), (ii)
confusion limited maps at $\sim 2$~mJy (which means that only the
brightest members are detected) and (iii) insufficient resolution
($\sim 15\arcsec$) to resolve extended lensed structures. Current
efforts to increase the surveyed area at 850~$\mu$m include the
SASS\footnote{SCUBA-2 All Sky Survey, a $\sim 2 \times 10^4$ square
degree survey with a $5 \sigma$ depth of 150 mJy and $~15\arcsec$
resolution}, the SCLS\footnote{SCUBA-2 Cosmology Legacy Survey, a
$\sim 35$ square degree survey with a $5 \sigma$ depth of 3.5 mJy and
$~15\arcsec$ resolution} and the all sky survey that will be carried
out with the HFI bolometer on board of the \emph{Planck}
satellite\footnote{$5 \sigma$ depth of $\sim 350$ mJy and
$\sim50\arcsec$ resolution}, but their resolutions will still be
insufficient to identify lensed arcs. The only instrument that can
currently provide sub-arcsecond resolution in submm (at $890~\mu$m) is
the SMA\footnote{The Submillimeter Array in Hawaii}.

However, the tight correlation between radio synchrotron and FIR
emission observed in star-forming galaxies \citep{VA73.1,HE85.1},
provides an alternative way to obtain high resolution images of SMGs.
Commonly referred to as "the FIR/radio correlation", it covers about
five orders of magnitude in luminosity \citep{CO92.1,GA02.1} out to
$z\sim 3$ \citep{KO06.1,VL07.1,IB08.1,MI09.1}. The standard model
about its nature considers that both emissions are caused by massive
star formation: While young massive stars produce UV radiation that is
re-emitted in the FIR by the surrounding dust, old massive stars
explode as supernovae, producing electrons that are accelerated by the
galactic magnetic field and generate the observed radio synchrotron
emission \citep{HA75.1,HE85.1}. Therefore, given the possible common
physical origin of both emissions, radio interferometric observations
can be used as a high-resolution proxy for the rest-frame FIR emission
of high-$z$ galaxies observed in the submm.

The advent of ALMA\footnote{The Atacama Large Millimeter Array in
Chile} will open a new window into mm/submm astronomy at sub-arcsecond
resolution and sub-mJy sensitivity, allowing the detection of resolved
gravitational arcs produced by SMGs. Although its small instantaneous
field of view (FOV) severely limits ALMA's survey capability, a 25
meter submm telescope (CCAT\footnote{Cornell Caltech Atacama
Telescope}) is going to be built on a high peak in the Atacama
region to provide wide field images ($\sim 400$ arcmin$^2$)
with a resolution of $\sim 3.5\arcsec$ at $350~\mu$m. With the
combined capabilities of both instruments, arc statistics studies in
the submm might be possible.

At the same time, radio interferometry is also experiencing
major technological improvements. In particular, the
VLA\footnote{The Very Large Array in New Mexico} and
MERLIN\footnote{The UK Multi-Element Radio Linked Interferometer
Network} are currently undergoing major upgrades which will boost
their sensitivities by factors of $10-30$ and dramatically improve their
mapping capabilities. The new versions of these arrays
(\emph{e}-MERLIN and EVLA) will be fully operational in 2010 and 2012
respectively. In order to assess the prospects for the study of
gravitationally lensed arcs at submm and radio wavelengths, in this
work we report detailed theoretical predictions about the abundance of
arcs produced by the SMG population at $850~\mu$m and 1.4~GHz.

The paper is organized as follows. In Section
\ref{sct:statistics}, we introduce all the relevant quantities that
are necessary to derive the total number of arcs detectable on the
sky. In Section \ref{sct:general} we present and discuss all the
observational information about SMGs that is required for the
subsequent strong lensing analysis: morphology, redshift distribution
and number counts. A description of our cluster population model and
the way in which the abundance of large arcs is computed is given in
Section \ref{sct:model}. The derived arc redshift distributions and
number of arcs are presented in Section \ref{sct:results}, while in
Section \ref{sct:relations} we discuss how these relate to previous
findings in the literature. A summary and conclusions are presented in
Section \ref{sct:conclusions}.

The adopted cosmology corresponds to the standard $\Lambda$CDM model with cosmological parameters inferred from the WMAP-$5$ data release in conjunction with type Ia supernovae and baryon acoustic oscillation datasets \citep{DU09.1,KO09.1}, namely $\Omega_{\mathrm{m},0} = 0.279$, $\Omega_{\Lambda,0} = 0.721$, $\sigma_8 = 0.817$ and $H_0 = h 100$ km s$^{-1}$ Mpc$^{-1}$,
with $h = 0.701$.

\section{Strong lensing statistics}\label{sct:statistics}

The choice of the best parameters to be used in order to characterize
the morphological properties of long and thin gravitational arcs is
still a matter of debate. In this work, we adopted the quite popular
choice of the length-to-width radio $d$, which has to be
larger than a certain threshold $d_{0}$ (usually $7.5$ or $10$) in
order to consider an object as a giant arc.

Given a set of background sources placed at redshift $z_\mathrm{s}$,
the efficiency of the galaxy cluster population to produce arcs with
length-to-width ratio $d \ge d_0$, is parametrized by the optical
depth

\begin{equation}\label{eqn:tau}
\tau_{d_0}(z_\mathrm{s}) = \frac{1}{4\pi D_\mathrm{s}^2} \int_0^{z_\mathrm{s}}\int_0^{+\infty} n(M,z)\sigma_{d_0}(M,z)dM dz,
\end{equation}
where $D_\mathrm{s}$ is the angular diameter distance to the source
redshift and $n(M,z)$ is the total number of clusters present in the
unit redshift around $z$ with mass in the unit interval around
$M$. The cross section $\sigma_{d_0}(M,z)$ is the area of the region
on the source plane where a source has to lie in order to produce (at
least) one gravitational arc with $d \ge d_0$, for a single cluster
with mass $M$ at redshift $z$. This depends in general on the cluster
structure, the source properties, and the redshifts of both the
cluster and the source. Since in realistic situations sources are
distributed at different redshifts, we can calculate the average
optical depth by integrating $\tau_{d_0}(z_\mathrm{s})$ over the
source redshift distribution $p(z_\mathrm{s})$,

\begin{equation}\label{eqn:av}
\bar{\tau}_{d_0}(z_\mathrm{s}) = \int_0^{z_\mathrm{s}}p(\xi)\tau_{d_0}(\xi)d\xi.
\end{equation}

In this way, the total number of arcs with length-to-width
ratio $d \ge d_0$ can be calculated as \citep{BA98.1}

\begin{equation}
\mathcal{N}_{d_{0}} = 4\pi N~\bar{\tau}_{d_0},
\end{equation}
where $N$ is the observed surface density of sources, and
$\bar{\tau}_{d_0} = \bar{\tau}_{d_0}(+\infty)$ is the total average
optical depth (i.e., the average optical depth with the integral extending to
all possible source redshifts). Therefore, the number of arcs produced by a given population of
background sources can be calculated by providing the following
observational constraints: (i) the characteristic source shape and
size, which is necessary to compute the cluster cross sections
$\sigma_{d_0}(M,z)$, (ii) the source redshift distribution $p(z_{s})$,
which is necessary to evaluate the optical depth $\bar{\tau}_{d_0}$,
and (iii) the cumulative source number counts $N$.

\section{Characteristics of the SMG population}\label{sct:general}

\subsection{Source shape and size}\label{sct:shape}

In strong lensing statistics studies, it is customary to characterize
the size of elliptical background sources using the equivalent
effective radius $R_\mathrm{e} \equiv \sqrt{ab}$, which is the radius
of a circle that has the same area of the elliptical source with
semi-major axis $a$ and semi-minor axis $b$. In addition, the
orientation of sources is randomly chosen, and to account for the
different source shapes the value of the axis ratio $b/a$ is
considered to vary within a certain interval. The typical values of
these parameters used in optical ray-tracing simulations are
$R_\mathrm{e} = 0.5\arcsec$ and $b/a$ randomly varying in the interval
$[0.5,1]$ \citep{ME00.1,ME03.1,ME05.1,FE06.1}.

Due to the low resolution of submm single-dish observations (e.g.,
FWHM $\sim 15\arcsec$ for SCUBA at $850\mu$m), current estimates of
the typical size of SMGs are based on continuum radio
\citep{CH04.1,BI08.1} and millimeter \citep[e.g.][]{TA06.1}
interferometric observations of small source samples. In particular,
Biggs \& Ivison (BI08 hereafter) combined $1.4$~GHz data from the VLA
and MERLIN to produce high resolution radio maps of $12$ SMGs detected
in the Lockman Hole. The deconvolved sizes derived from these radio
maps (obtained by fitting each map with an elliptical Gaussian) are
consistent with \cite{CH04.1} (see also \citealt{MU05.1}) and
\cite{TA06.1}.

\begin{figure}[t]
  \includegraphics[width=\hsize]{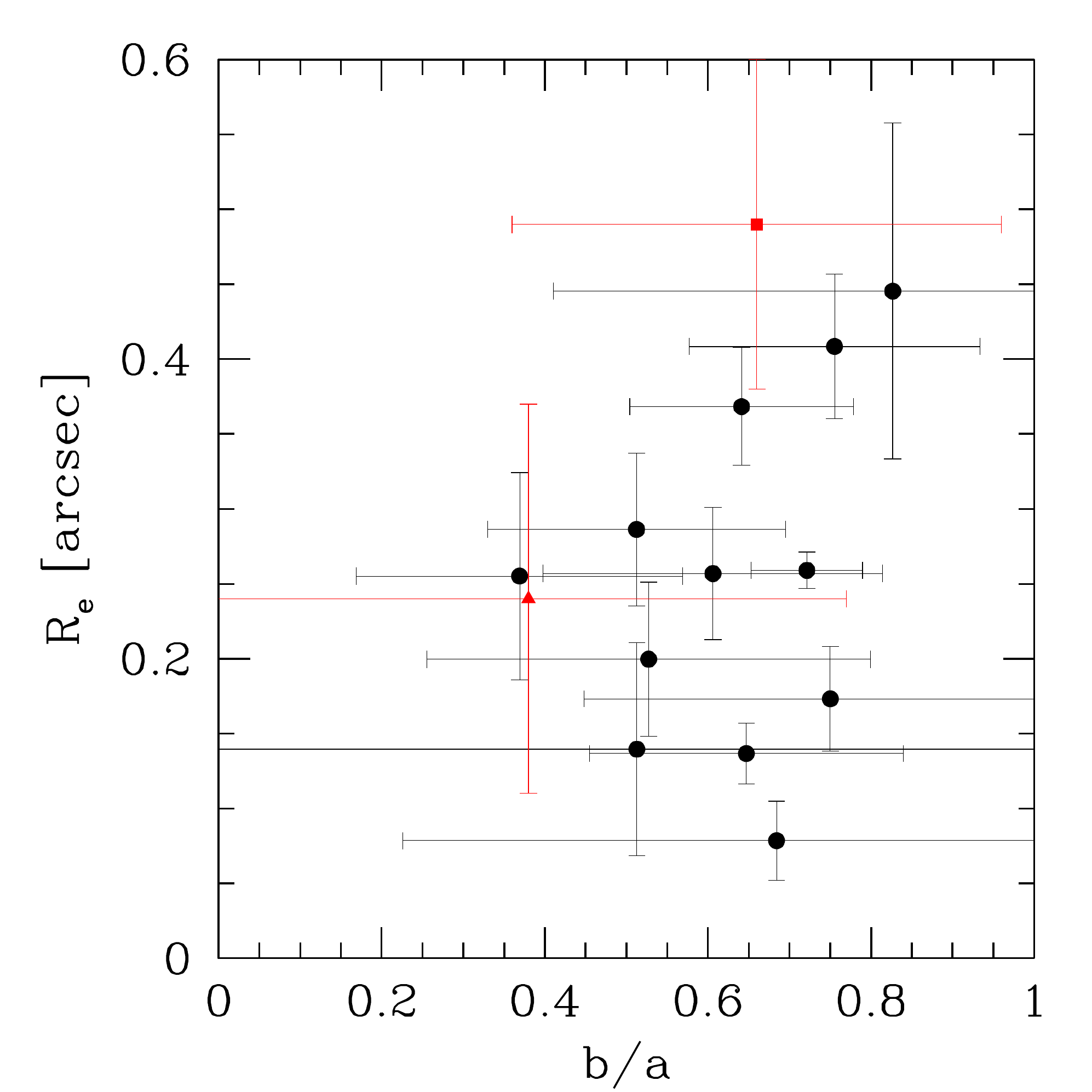}\hfill
\caption{Effective radii and axis ratios derived for the $1.4$~GHz
radio counterparts of the $12$ SMGs presented in \cite{BI08.1}.  The red
points correspond to the $890~\mu$m counterpart of GN20 detected with
the SMA, whose size was derived by fitting a Gaussian (triangle) and
an elliptical disk (square) to the data \citep{YO08.1}. Error bars
are computed through standard error propagation.}
\label{fig:ad}
\end{figure}

\begin{figure*}[ht!]
  \includegraphics[width=0.5\hsize]{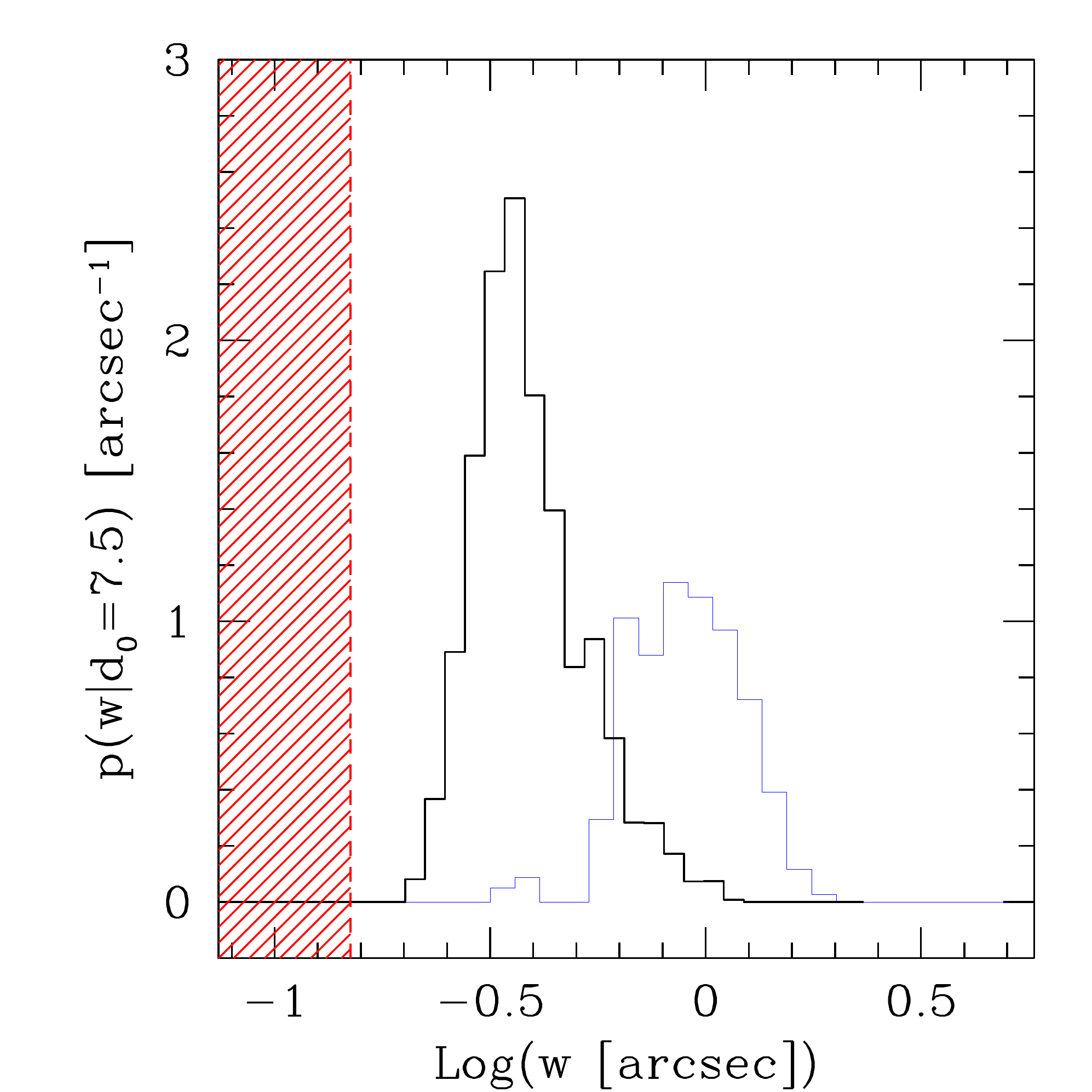}\hfill
  \includegraphics[width=0.5\hsize]{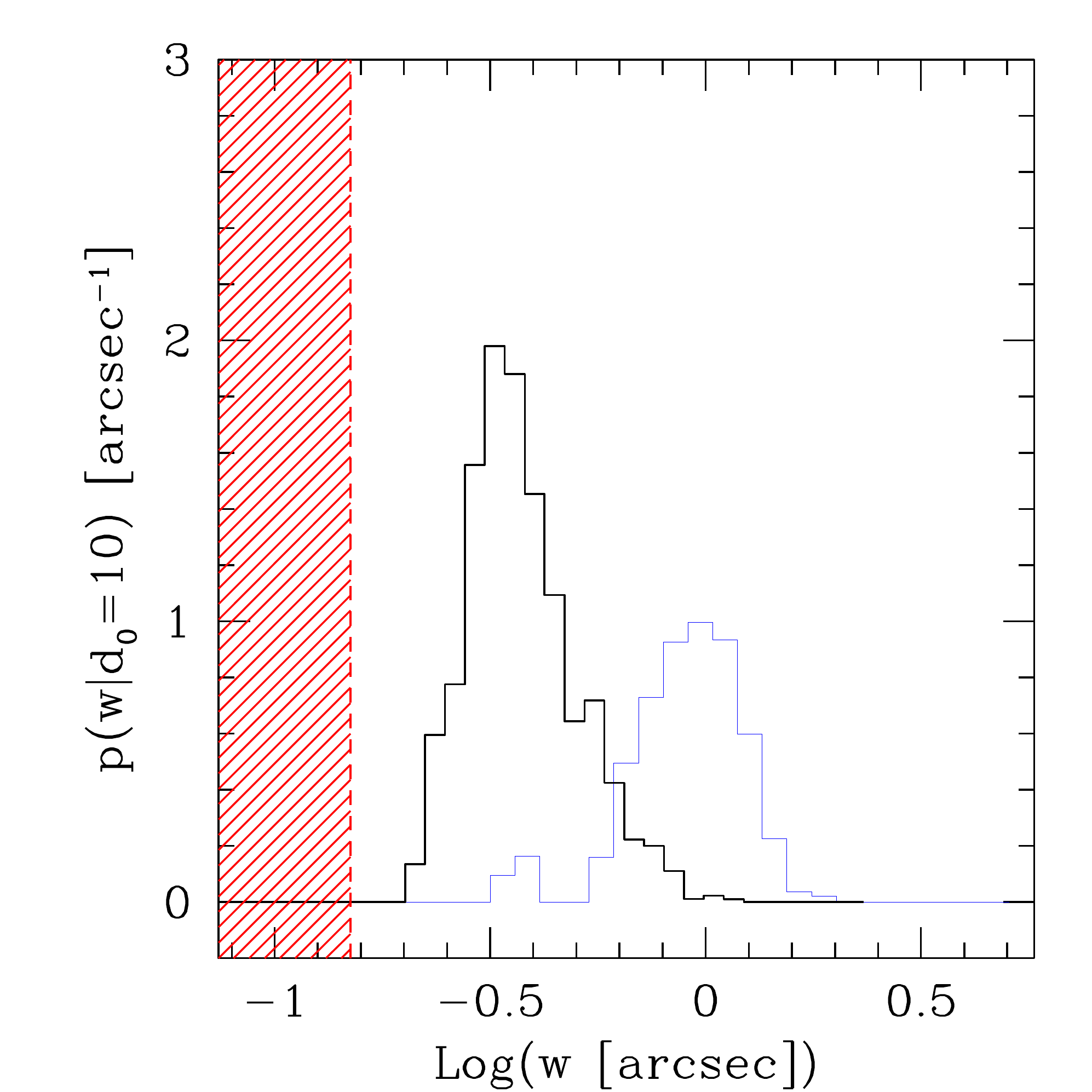}
\caption{Arc width probability distributions derived from a set of
sources at $z_\mathrm{s}=1$ that were lensed by (the three projections of)
the most massive cluster of the \emph{MareNostrum} simulation
at $z=0.3$. The two panels illustrate the difference between selecting
arcs with a length-to-width ratio $d \ge d_0 = 7.5$ (left) and $d \ge
d_0 = 10$ (right). The black histograms were produced using the
typical source size of SMGs at 1.4~GHz and $850~\mu$m assumed in this
work ($R_\mathrm{e} = 0.25\arcsec$ and $b/a$ randomly varying in the
interval $[0.3,1]$). The blue histograms correspond to optical sources
as usually described in lensing simulations ($R_\mathrm{e} =
0.5\arcsec$ and $b/a \in [0.5,1]$). The red dashed region shows the
$150$ mas resolution limit of the $e$-MERLIN radio interferometer at
$1.4$~GHz.}
\label{fig:wd}
\end{figure*}

Figure \ref{fig:ad} shows the effective radius and axis ratio for each
of the $12$ radio sources reported in BI08 (black circles). Note that,
although the $b/a$ interval $[0.5,1]$ used in optical lensing
simulations contains 11 out of the 12 radio sources, there are several
error bars that extend below its lower limit. In addition, the optical
effective radius of $0.5\arcsec$ is not suitable to describe
them. Since the source sample is too small (and the error bars rather
large) to derive a reliable size distribution , we decided to use an
effective radius close to the median of the sample. As a result, the
size of the 1.4~GHz radio emission produced by the SMG population has
been characterized in the following by $b/a$ randomly varying within
$[0.3,1]$ and $R_\mathrm{e} = 0.25\arcsec$.

Since SMGs seem to follow the FIR-radio correlation
\citep[e.g.][]{KO06.1}, their emission at both 1.4~GHz and $850~\mu$m
is expected to be associated with massive star formation.  This means
that, as a first approximation, the same morphological parameters can
be used to characterize the sizes and shapes of SMGs at radio and
submm wavelengths. This choice of parameters for the submm emission is
also consistent with the recent SMA observations presented by
\cite{YO08.1}, which have partially resolved the $890~\mu$m emission
of a SMG (GN20) for the first time (see Figure \ref{fig:ad}).

If we wish to characterize gravitational arcs via their
length-to-width ratio and make comparisons between observations and
theoretical predictions in an unbiased way, it is crucial to resolve
their width. To address if the resolution provided by radio and submm
instruments could be an issue for arc statistics studies, we investigated the width distribution of arcs produced by a population
of sources that is being lensed by a galaxy cluster. In particular, we
used the most massive lensing cluster at $z=0.3$ from the
\emph{MareNostrum} cosmological simulation \citep{GO07.1}, a large
$n$-body and gas-dynamical run, whose lensing properties recently have been studied by Fedeli et al. (2009, in preparation). The mass
distribution of this cluster was projected along three orthogonal
directions, for which we derived deflection angle maps by standard
ray-tracing techniques \citep{BA94.1}. Then, a set of sources at
$z_\mathrm{s} = 1$ (source redshift at which the lensing efficiency
peaks for lenses at $z \sim 0.3$) with $R_\mathrm{e} = 0.25\arcsec$
and axis ratios randomly varying in the interval $[0.3,1]$ was lensed
through the three projections. As usual in this procedure, the sources
are preferentially placed near the lensing caustics following an
iterative procedure to enhance the probability of the production of
large arcs. The bias introduced by this artificial increase of sources
is corrected for by assigning a weight $\le 1$ to each source, which is
reduced at each new iteration step (see \citealt{MI93.1,BA94.1,BA98.1}
for further details).

The black histogram shown in Figure \ref{fig:wd} corresponds to the
width distribution of arcs derived from this simulation. For
comparison, we have also included the corresponding distribution
derived from the parameters used in optical lensing simulations (blue
histogram). The two panels illustrate the difference between selecting
arcs characterized by $d \ge d_0$ with $d_0 = 7.5$ (left) and $d_0 =
10$ (right). As expected, reducing the source equivalent size produces
a decrease in the width of lensed images. Note also that the behavior
of the distributions is practically independent of the minimum
length-to-width ratio used to select the arcs. The most important
feature, however, is that both distributions drop to zero for widths
below $\sim 0.2\arcsec$, meaning that virtually no radio/submm (or
optical) arcs have widths smaller than that value. At 1.4~GHz, $\sim
0.2\arcsec$ resolution is already accessible with
MERLIN/\emph{e}-MERLIN ($\sim0.15\arcsec$). Therefore, resolving the
width of long and thin images for arc statistics studies is in
principle already possible at radio wavelengths. However, until the
advent of ALMA, the $\sim 0.75\arcsec$ resolution of the SMA at
$950~\mu$m will only be able to resolve a very small fraction of
arcs produced by the most extended SMGs.

\begin{figure*}[ht!]
  \includegraphics[width=0.5\hsize]{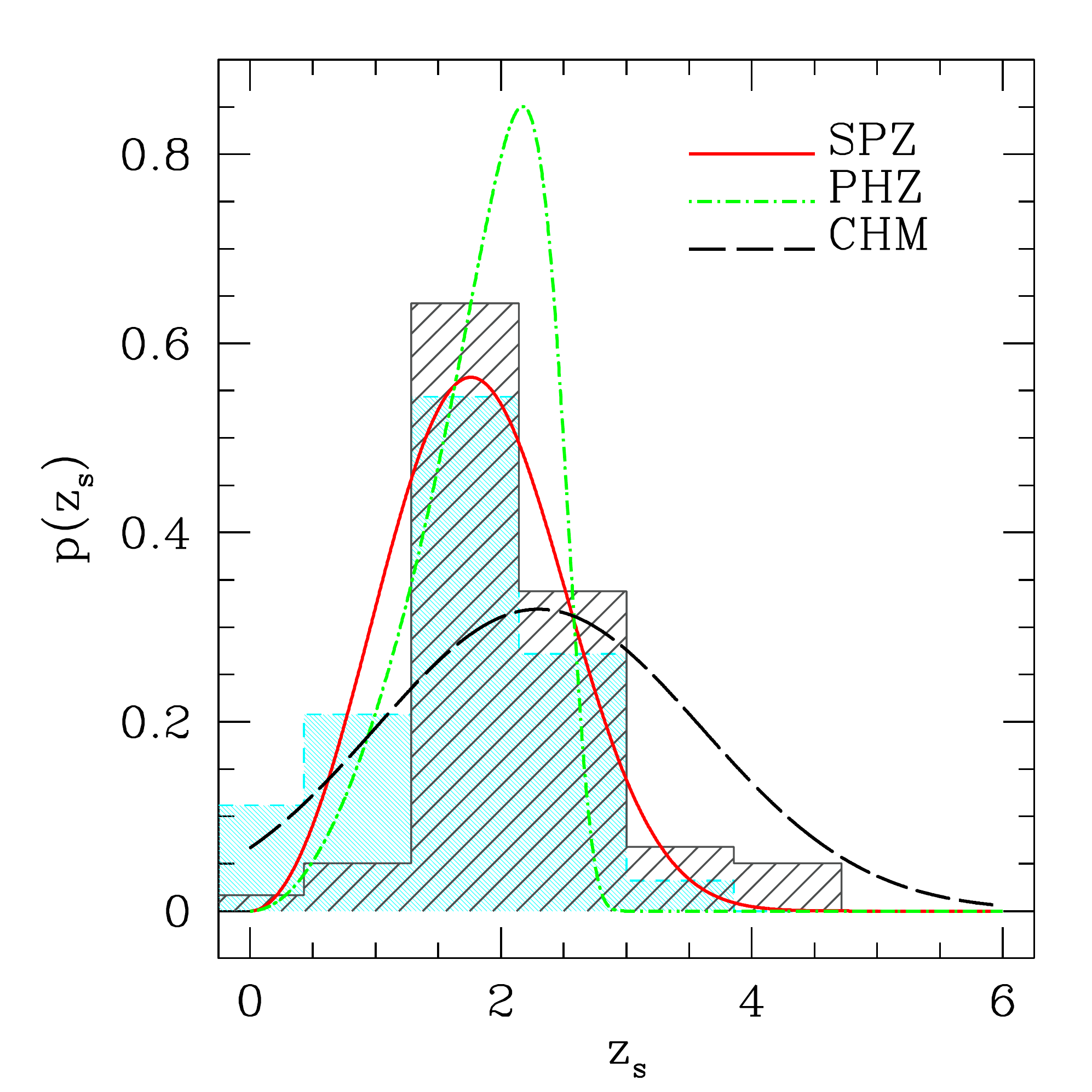}\hfill
  \includegraphics[width=0.5\hsize]{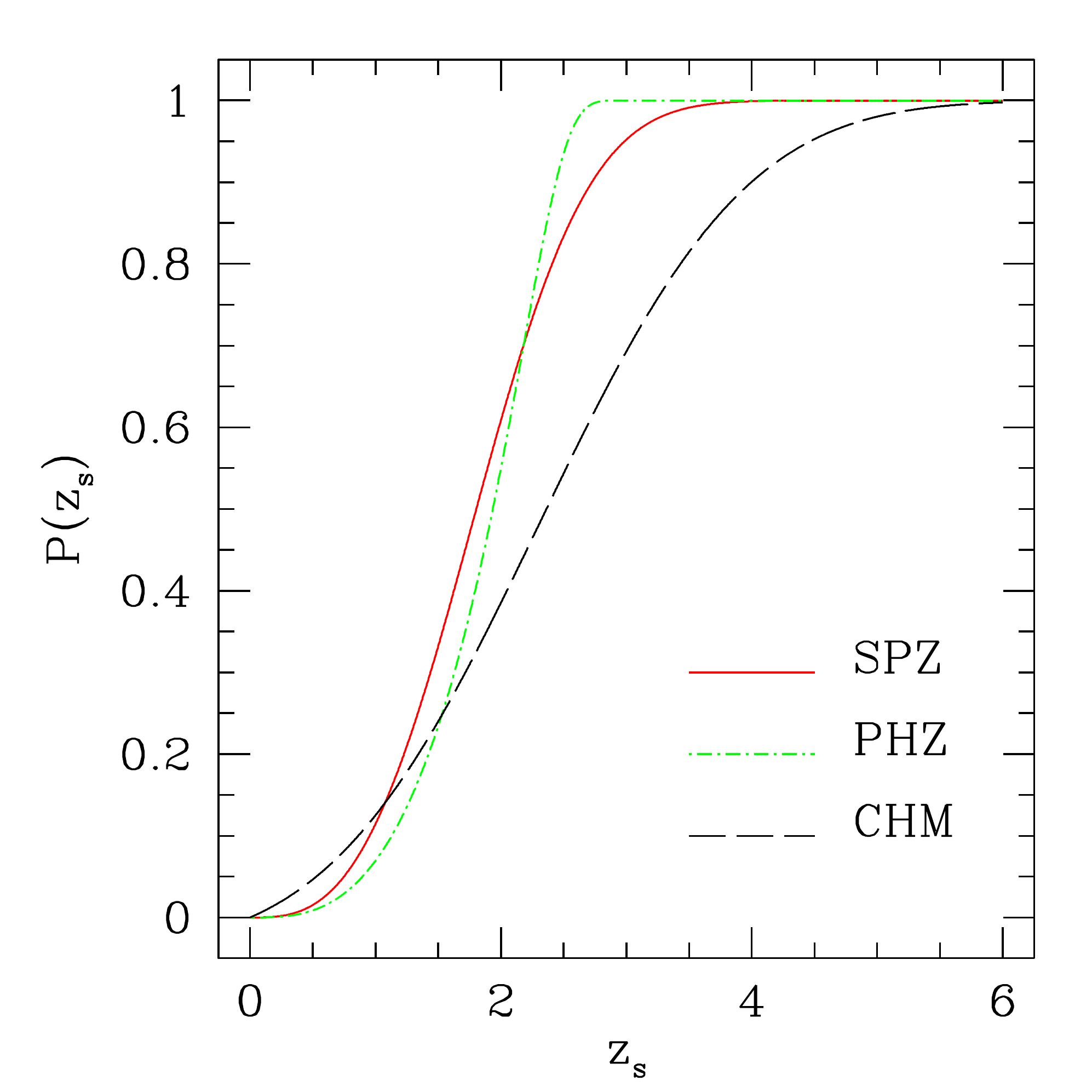}
\caption{\emph{Left panel}. Redshift distribution of SMGs derived
from the spectroscopic sample of \cite{CH05.1} (cyan histogram) and
the photometric sample of \cite{AR07.1} (dark-grey histogram). The curves
SPZ and PHZ correspond to the best fits provided by Eq. (\ref{eqn:zd})
to the spectroscopic and photometric data, respectively. The curve CHM
corresponds to the best Gaussian fit to the simple evolutionary model
for SMGs used in CH05 (as quoted in CH05), normalized to the redshift
interval $[0,+\infty]$. Note that the binning adopted here is
different from the one used in CH05, so the "redshift
desert" in the redshift interval $z=1.2-1.8$ is not evident. However, changing
the binning of the histogram changes little of the subsequent
fit. \emph{Right panel}. The cumulative distributions corresponding
to the fits reported on the left panel, calculated by using
Eq. (\ref{eqn:pz_cum}).}
\label{fig:zd}
\end{figure*}

Finally, we would like to stress two points related to the choice of
source morphological parameters presented in this section. First, the
most luminous SMGs seem to be the result of merger processes (e.g.,
\citealt{GR05.1,TA06.1}), hence it is unlikely that their true shape
is elliptical, as we assumed. However, if a merging source is lensed
as an arc at a particular wavelength, irregularities in its shape and
internal structure will not significantly change the global
morphological properties of the arc, like the length-to-width
ratio. What can happen is that the length-to-width ratio of an arc
changes with wavelength because the emitting region of the source at
those wavelengths have different sizes. For some of the wavelengths
the source might even look like a group of small isolated emitting
regions instead of a continuous one, which means that in the image
plane it will be observed as a group of disconnected multiple images
instead of a full arc. A very illustrative example of this scenario is
SMM J04542-0301, an elongated region of submm emission which seems to
be associated with a merger at $z=2.9$ that is being lensed by the
cluster MS0451.6-0305 (\citealt{BO03.1,BE07.1}; Berciano Alba et
al. 2009, in press). Until more complete information about
the average structure of SMGs becomes available, we believe our
approach to be the best that can be done.

Second, since the radio sources studied by BI08 are brighter than
$50~\mu$Jy, the typical size derived from this sample might be
different from the one that could be derived from fainter SMGs. Note,
however, that the change of the cluster cross section with source size
has a very small slope for $R_\mathrm{e}$ between 0.2$\arcsec$ and 1.5$\arcsec$
\citep{FE06.1}. Therefore, deviations from $R_\mathrm{e}=0.25\arcsec$ within this
interval (which is two times larger than the interval that contains
the sizes measured by BI08 and \cite{TA06.1}, see Figure 6 of BI08) will
have a negligible effect on the derived number of arcs.

\subsection{Source redshift distribution}\label{sct:redshift}

A key point in trying to estimate the abundance of strong lensing
features that are produced by the galaxy cluster population is the
redshift distribution of background sources. Distributions peaked at
higher redshift, or with a substantial high-$z$ tail, will have in
general more potential lenses at their disposal, and hence will produce
larger arc abundances as compared to low-$z$-dominated
distributions. In addition, the lensing efficiency for individual
deflectors is also an increasing function of the source redshift.

The most robust estimate of the redshift distribution of SMGs to date
is based on the $\sim$ 15~arcmin$^2$ SCUBA survey carried out by
\cite{CH05.1} (CH05 hereafter). Radio observations were used to
pinpoint the precise location of the submm detections, allowing the
identification of optical counterparts that could provide precise
spectroscopic redshifts. The final sample is composed of $73$ SMGs
with $850~\mu$m flux densities $>3$~mJy and radio counterparts with
flux densities at $1.4$~GHz $>30~\mu$Jy.

On the other hand, the SCUBA Half-Degree Extragalactic Survey (SHADES,
\citealt{MO05.1,VA05.1}) is the largest ($720$ arcmin$^2$) $850~\mu$m
survey to date\footnote{The largest SMG survey to date (0.7
deg$^{2}$) has being carried out at $1.1$ mm with the AzTEC continuum
camera mounted on the JCMT \citep{AU09.1}}. From their catalog of 120
SMGs, 69 have robust radio counterparts with $S_{850~\mu m} \geq$
3~mJy and $S_{1.4~\mathrm{GHz}} \geq$ 20~$\mu$Jy. Photometric
redshifts for this sub-sample were calculated by \cite{AR07.1} (AR07
hereafter) by fitting Spectral Energy Distribution (SED) templates to
the available photometry at $850~\mu$m and $1.4$~GHz, including upper
limits at $450~\mu$m (additional photometry at millimeter wavelengths
was also used for $13$ out of the $69$ sources). The histogram of the
resulting photometric redshift distribution, together with the
spectroscopic one reported by CH05, are shown in Figure
\ref{fig:zd}. The accuracy on the photometric
redshifts derived by AR07 is $\Delta z \sim 0.65$. Note
that the requirement for a radio counterpart biases these two redshift
distributions against SMGs with $z > 3$, due to the less favorable
K-correction in the radio compared with submm. Using a simple
evolutionary model, CH05 estimated that the fraction of SMGs
($S_{850~\mu m} >$ 5~mJy) that is missing between $z \sim 2.5$ and $z
\sim 5$ in their spectroscopic redshift distribution is $\sim 35\%$, a
number that is in agreement with the fraction of radio-unidentified
SMGs reported in \cite{IV02.1,CH03.1} and the SHADES survey
(AR07). The model evolves the local FIR luminosity function in
luminosity with increasing redshift following the prescription of
\cite{BL02.1}. To account for the dust properties of SMGs, it also
includes a range of SED templates that have been tuned to fit their
observed submm flux distribution \citep{CH03.1,LE05.1}.

\begin{figure*}[th!]
  \includegraphics[width=0.5\hsize]{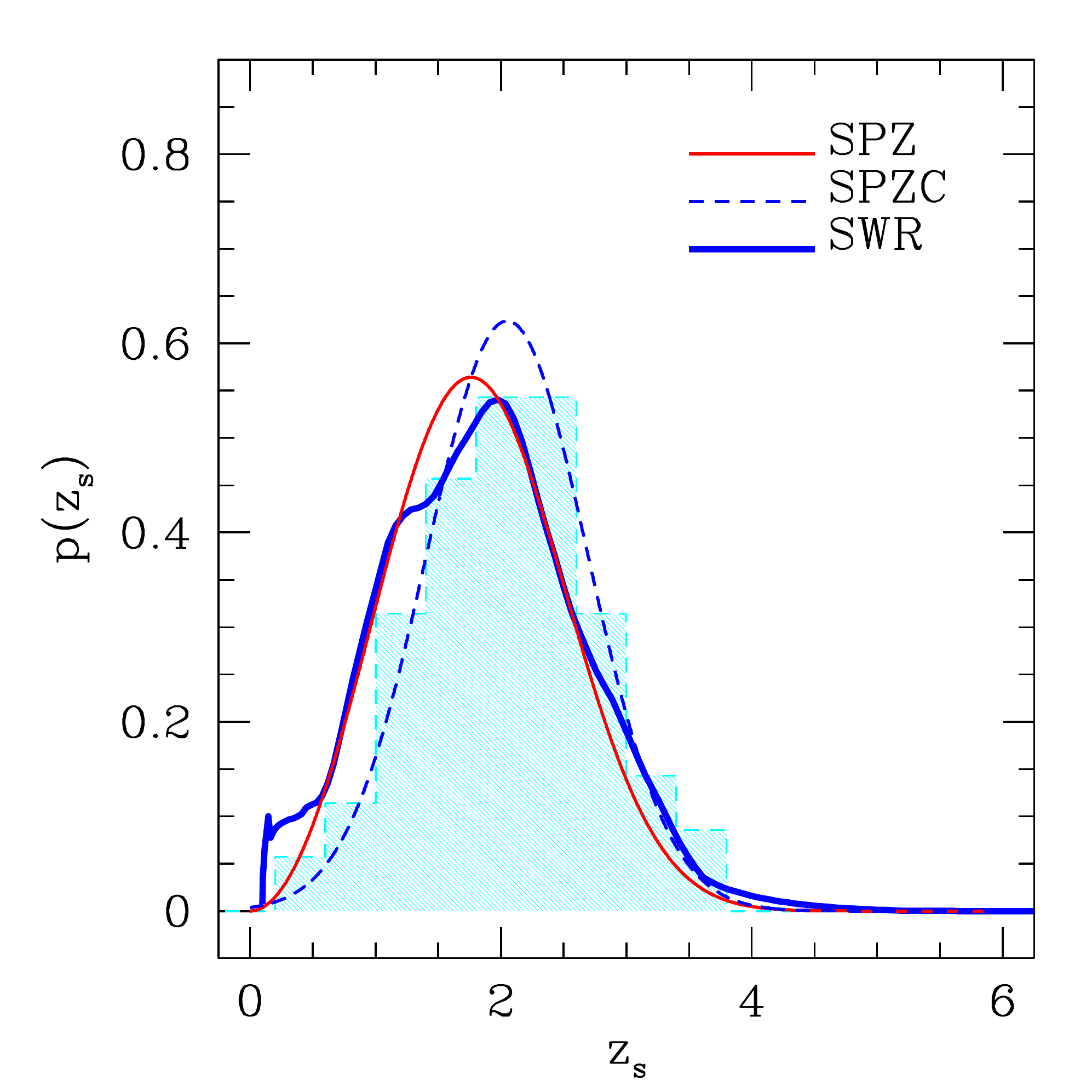}\hfill
  \includegraphics[width=0.5\hsize]{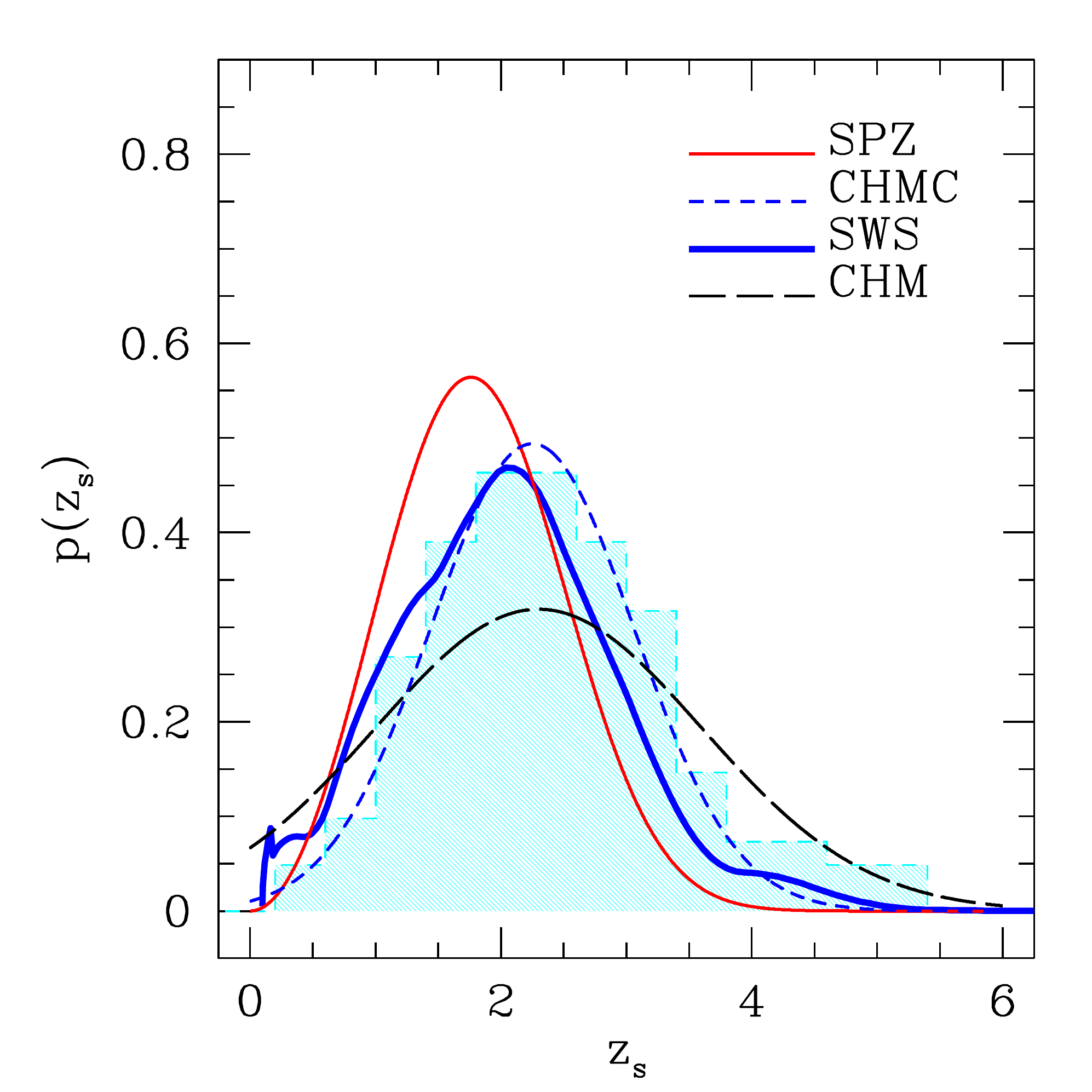}
\caption{\emph{Left panel}. Histogram of the redshift distribution
of SMGs derived by \cite{CH05.1}, corrected for spectroscopic
incompleteness. This correction was implemented by interpolating the CH05
distribution in the region of the redshift desert (M.
Swinbank, private communication). The SPZC line indicates the best
Gaussian fit to the histogram. The blue solid line corresponds to the
redshift distribution of SMGs with $S_{850~\mu m} >$ 5~mJy and $S_{\rm
1.4~GHz} >$ 30~$\mu$Jy predicted by the semi-analytic model presented
in \cite{SW08.1}. The SPZ curve presented in Figure \ref{fig:zd} has
being included for comparison. \emph{Right panel}. The same
histogram, also corrected for radio incompleteness using CH05
evolutionary model. The blue solid line corresponds to the redshift
distribution of SMGs with $S_{850~\mu m} >$ 5~mJy predicted in
\cite{SW08.1}. The CHMC line indicates the best Gaussian fit to the
histogram. The SPZ and CHM curves presented in Figure \ref{fig:zd} have
being included for comparison.}
\label{fig:z_comparison}
\end{figure*}

In order to investigate the effect of this high-$z$ tail on the
predicted number of arcs, we used three different analytic expressions
to characterize the redshift distribution of SMGs in our calculations
(see Figure \ref{fig:zd}). One of them (CHM) corresponds to the best
Gaussian fit to the distribution predicted by the CH05 evolutionary model,
as quoted in CH05. The other two (SPZ and PHZ) were obtained by
fitting the CH05 and AR07 histograms with the following analytic
function, usually adopted for optical strong lensing studies
\citep{SM95.1},

\begin{equation}\label{eqn:zd}
p(z_\mathrm{s}) = \frac{\beta}{z_0^3 \Gamma(3/\beta)} z_\mathrm{s}^2 \exp \left[ -\left( \frac{z_\mathrm{s}}{z_0} \right)^\beta \right],
\end{equation}
where $\Gamma(x)$ is the complete Euler-gamma function evaluated at
$x$, $z_0$ is a free parameter that broadly selects the position of
the peak, and $\beta$ is another free parameter that defines the
extension of the high-redshift tail. Note that, unlike the histograms, this function drops to zero at low redshift. However,
the contribution to the global lensing optical depth coming from
sources at $z \lesssim 1$ is likely negligible, due to the geometric
suppression of lensing efficiency (see also the subsequent discussion
in Section \ref{sct:results}). The resultant best-fitting parameters
of these three functions are summarized in Table \ref{tab:par}.

\begin{table}[]
  \caption{Parameters of the redshift distributions presented in
  Figures \ref{fig:zd} and \ref{fig:z_comparison}.}

  \label{tab:par}
  \begin{center}
    \begin{tabular}{clcccc}
      \hline\hline

      \noalign{\smallskip}
      Nickname & $z_0^{(1)}$ & $\beta^{(1)}$ & \emph{rms$^{(2)}$} & $z_\mathrm{p}^{(3)}$ \\
      \noalign{\smallskip}
      \hline
      \noalign{\smallskip}
      SPZ       &  $1.99$ & $2.81$ &  $-$    & $1.76$ \\
      PHZ       &  $2.51$ & $13.3$ &  $-$    & $2.18$ \\
      CHM       &  $-$    & $-$    & $1.30$   & $2.30$ \\
      \noalign{\smallskip}
      \hline
      \noalign{\smallskip}
      SPZC$^{(4)}$  & $-$     & $-$    & $0.64$     & $2.05$  \\
      CHMC$^{(4)}$  & $-$     & $-$    & $0.81$     & $2.25$  \\
      \hline
\noalign{\smallskip}
\multicolumn{5}{l}{$^{(1)}$ best fit parameters of Eq. (\ref{eqn:zd}).}\\
\multicolumn{5}{l}{$^{(2)}$ width of the best Gaussian fit.}\\
\multicolumn{5}{l}{$^{(3)}$ peak position of the distribution.}\\
\multicolumn{5}{l}{$^{(4)}$ the parameters associated with these}\\
\multicolumn{5}{l}{distributions, provided by M. Swinbank,}\\
\multicolumn{5}{l}{were misquoted in SW08 and CH05.}\\

    \end{tabular}
  \end{center}
\end{table}

The SPZ curve constitutes a good representation of CH05 data not
corrected for spectroscopic incompleteness\footnote{Due to the lack of
strong spectral features falling into the observational windows, it is
not possible to measure the spectroscopic $z$ of sources in the range
$z=1.2-1.8$ (the so called "redshift desert")}. The PHZ curve, on the
other hand, does not describe the AR07 distribution very well, failing
to reproduce its $z\sim3-4.5$ tail. This is a consequence of how the
function in Eq. (\ref{eqn:zd}) is constructed. In particular, to
accommodate the low-$z$ part of the photometric histogram in Figure
\ref{fig:zd}, the function needs to raise very steeply and hence, by
construction, it must also drop steeply at high-$z$. Despite the function
in Eq. (\ref{eqn:zd}) not being a good choice for fitting the
photometric data, we nevertheless included the PHZ curve in our
calculations because it highlights the consequence of choosing a
distribution that is truncated at $z\sim3$. Finally, the CHM curve
allows us to predict the number of expected arcs if the CH05 histogram is
corrected for spectroscopic incompleteness and high-z SMGs without
radio counterparts (radio incompleteness). We stress that at this
stage we are not interested in using the best possible representation
for the true source redshift distribution, but only to adopt a few
motivated choices that broadly cover the range of realistic
possibilities, in order to check the corresponding effect on the
abundance of arcs.

To show more clearly the different behavior in the high-redshift tail of our three choices, we
present their cumulative distributions in the right panel of
Figure \ref{fig:zd}, namely

\begin{equation} \label{eqn:pz_cum}
P(z) = \int_0^z p(\xi)d\xi.
\end{equation}
In particular, when $P(z) \simeq 1$ for SPZ, at $z \sim 3.5$, we still
have $P(z) \simeq 0.8$ for CHM, implying that $\sim 20\%$ of the
SMGs still can be found at $z \gtrsim 3.5$ using the latter distribution.

In order to further show that this family of three functions cover all
the reasonable possibilities, we have compared them with the
predictions of one of the semi-analytical models that have been
developed to explain the properties of SMGs (see \citealt{SW08.1} and
references therein). The histogram in the left panel of Figure
\ref{fig:z_comparison} shows CH05 data corrected for spectroscopic
incompleteness. Note that the SPZ curve is very consistent with the
semi-analytic model prediction (SWR), although both curves peak at
slightly lower redshift ($\Delta z_\mathrm{p}=0.3$) than the best
Gaussian fit of the histogram (SPZC). However, as pointed out in
\cite{SW08.1} (SW08 hereafter) , the CH05 distribution is expected to
be uncertain by at least $\Delta z \sim 0.25$, which is the
field-to-field variation between the seven sub-fields in the CH05
sample due to cosmic variance. Therefore, we can consider SPZ as a
good representation of the current observations, despite the fact that
it comes from a histogram that was not corrected for spectroscopic
incompleteness.

After the computations of the number of arcs were completed, we
became aware of the fact that the parameters quoted in CH05 for the
best Gaussian fit to their simple evolutionary model (CHM, see Table
\ref{tab:par}) were incorrect (M. Swinbank, private
communication). As it is shown in the right panel of Figure
\ref{fig:z_comparison}, the CHM distribution has a higher-$z$ tail as
compared to the correct Gaussian fit (CHMC) and the prediction of the
semi-analytical model (SWS). Since the true high-$z$ tail of the
redshift distribution of SMGs is expected to be in between the cases
considered in our calculations (SPZ, PHZ and CHM), and (as it will be
discussed in Section \ref{sect:arc_results}) the final effect of the
source redshift distribution on the number of arcs is small given the
many uncertainties involved, we considered it unnecessary to repeat the
calculations for CHMC.

\subsection{Source number counts}\label{sct:counts}

The final ingredient needed to estimate the number of arcs produced by
SMGs is the observed surface density of this source population, both
at $1.4$~GHz and $850~\mu$m. Let $n_0(S)$ be the differential number counts, defined as the surface
density of unlensed galaxies per unit flux density $S$.  Integrating
$n_0(S)$ over all fluxes above a given threshold, we obtain the
respective cumulative number counts

\begin{equation}\label{eqn:cum}
N_0(S) \equiv \int_S^{+\infty} n_0(\xi) d\xi.
\end{equation}
Let $\mu$ be the lensing-induced magnification of images on the
lens plane, and $\mu_+ \equiv |\mu|$. If $P(\mu_+|d_0)$ is the
magnification probability distribution for sources that are imaged
into arcs with $d \ge d_0$, then the magnified differential number
counts can be calculated as \citep{BA01.1}

\begin{equation}
n(S) = \int_0^{+\infty} n_0\left( \frac{S}{\mu_+} \right) \frac{P(\mu_+|d_0)}{\mu_+^2} d\mu_+.
\end{equation}
Hence, the magnified cumulative number counts read as

\begin{equation}\label{eqn:mag}
N(S) \equiv \int_S^{+\infty} n(\xi) d\xi = \int_0^{+\infty} N_0\left( \frac{S}{\mu_+} \right) \frac{P(\mu_+|d_0)}{\mu_+} d\mu_+.
\end{equation}
As can be seen in Eq. (\ref{eqn:mag}), the lensing magnification bias
has a twofold effect. On one side, sources that would be too faint to
be detected without the action of lensing are amplified, and hence the
respective images are brought above the detection threshold. On the
other side, the unit solid angle is stretched by lensing
magnification, implying that the number density of sources is
decreased. Which one of these two effects wins depends on the local
slope of the unmagnified cumulative number counts. In particular, if
$N_0(S) \propto S^{-\alpha}$ and $\alpha > 1$, the number density of
sources will be increased, while if $\alpha < 1$ it will be
decreased. It should be noted that, while for the unmagnified
counts $N_0(S)$ the flux density is derived by integrating the surface brightness over the
area of the source, for the magnified counts $N(S)$ the integral is performed over the area of the resulting arc.

Since our main motivation was to provide predictions for the abundance
of giant arcs to be detected in surveys carried out with future
instruments, we needed to provide the predicted number of arcs as a
function of the surface brightness, instead of flux density. The
reason is that we are working under the assumption that arcs are
resolved structures, and therefore they are observed as extended
objects. Under these circumstances, the flux integrated over the
resolution element of the instrument (seeing, PSF, pixel or beam) is
no longer the total flux of the source (as in the case of unresolved
sources), and it may therefore be below the limiting flux although the
arc as a whole is not. In other words, arc detectability under these
circumstances is not limited by the flux density but rather by the
surface brightness.

In order to take this into account, we had to convert the observed
number counts as a function of flux density into number counts as a
function of surface brightness. Assuming that the size of sources
is given by $R_\mathrm{e}$, and that the surface brightness is
constant across it, then $N_0(B) = N_0(S/\pi R_\mathrm{e}^2)$. Note
that, since the surface brightness is not affected by lensing, the
magnification bias will manifest itself only through the solid angle
stretching. Therefore, the cumulative magnified number counts (as a
function of surface brightness) can be written as

\begin{equation}\label{eqn:mag_B}
N(B) = N_0(B)\int_0^{+\infty} \frac{P(\mu_+|d_0)}{\mu_+} d\mu_+.
\end{equation}
Among other things, this implies that the magnification bias will
always decrease the cumulative number counts, irrespective of the
shape of the unmagnified ones.

In the following we used the magnification distribution given by
\cite{FE08.1}, which is represented by the superposition of two
Gaussians. In particular, we adopted the $P(\mu_+|d_0)$ function for
$d_0 = 10$, but the result is virtually the same also for the case
$d_0 = 7.5$. Note however, that this (conditional) magnification
distribution was computed for a background population of sources that
have different morphologies than SMGs (see Section
\ref{sct:shape}). In principle, the bimodality of the magnification
distribution is expected to be preserved because it only depends on
the caustic structure \citep{LI05.1}, but it can be affected by the
source morphology in two opposite ways. On one hand, since SMGs are
smaller than in \cite{FE08.1}, we expect large arcs to form closer to
the critical curves, and therefore to have larger magnifications on
average. On the other hand, the fact that SMGs are more elongated will
favor the formation of large arcs in regions of lower
magnification. Given the uncertainties in other parts of the
calculation, we consider that the use of a magnification distribution
derived for optical sources will have a marginal effect on the derived
number of arcs produced by SMGs. For a comprehensive review of the
many effects that could affect the estimation of arc abundances by
galaxy clusters, see the discussion in \cite{FE08.1}.

\subsubsection{Submm number counts}

For the latest and most complete estimate of the submm number counts
at $850~\mu$m we refer to \cite{KN08.2} (hereafter KN08), who carried out a
combined analysis of the counts derived from the
Leiden SCUBA Lens Survey (LSLS) and the SHADES survey. With
an area of 720 arcmin$^2$, the SHADES survey is the largest
blank-field submm survey completed to date, and therefore the least
affected by cosmic variance. It provides the best constraints for the
submm number counts in the flux density range $2-15$ mJy
\citep{CO06.1}. On the other hand, the LSLS survey targeted
$12$ galaxy cluster fields which cover a total area of $71.5$
arcmin$^2$ in the image plane. It provides the deepest constraints at
the faint end of the submm counts ($0.11$~mJy, after correcting for
the lensing magnification). In their analysis, KN08 used two functions
to characterize the combined differential number counts from both
surveys: A double power-law,

\begin{equation}\label{eqn:dpl}
n_0(S) = \frac{n_{0,*}/S_*}{\left( S/S_* \right)^{\alpha} + \left( S/S_* \right)^{\beta}}
\end{equation}
and a Schechter function \citep{SC76.1},

\begin{equation}\label{eqn:schechter}
n_0(S) = n_{0,*}\left( \frac{S}{S_*} \right)^{\alpha+1} e^{-S/S_*}.
\end{equation}

\begin{figure}[]
  \includegraphics[width=1\hsize]{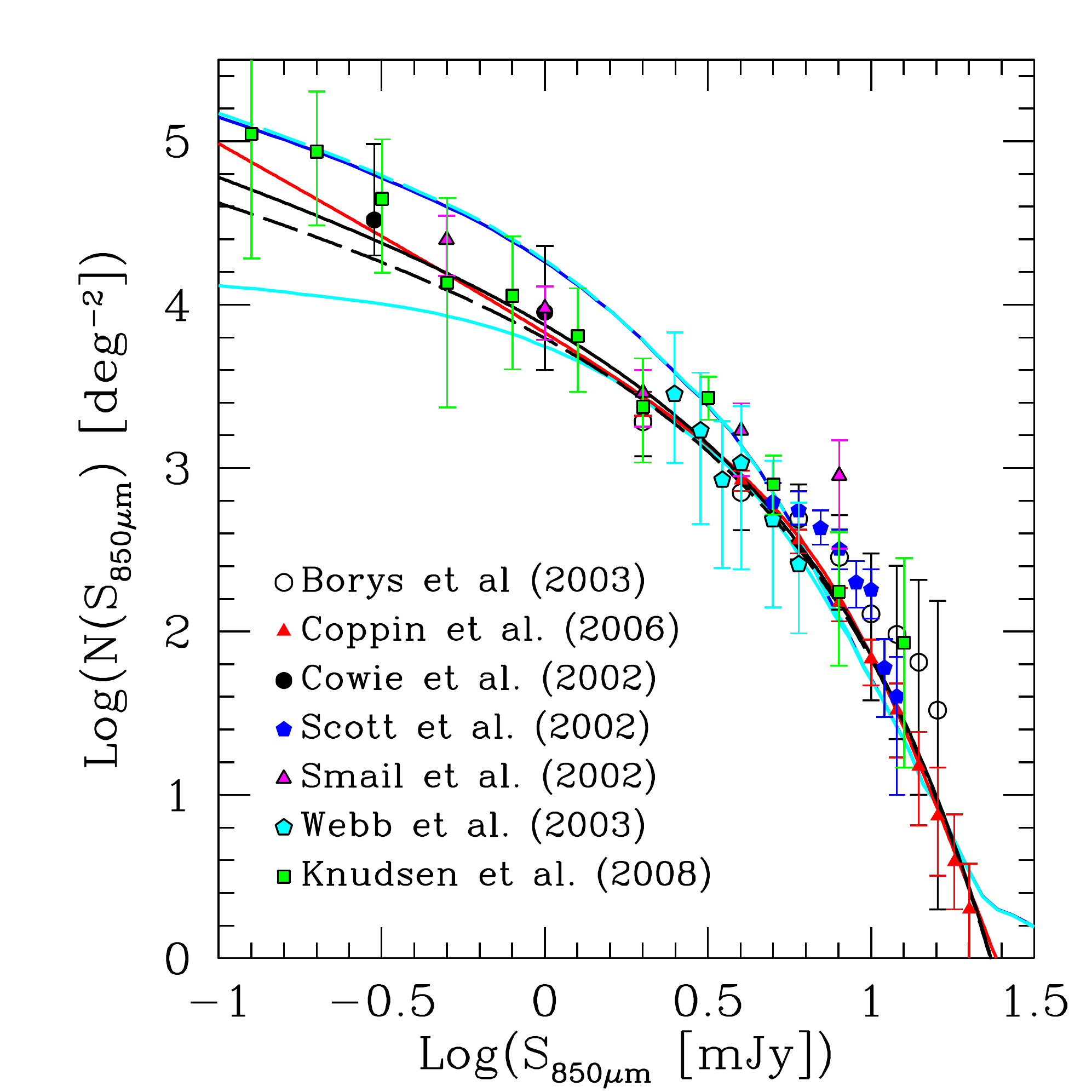}\hfill
\caption{Comparison between observed and predicted cumulative number
counts. The red and black solid curves corresponds to the
best-fit double power-law (DB) and the best-fit Schechter function
(SB) derived by \cite{KN08.2} for the combined $850~\mu$m cumulative
number counts from the LSLS survey \citep{KN08.2} and the SHADES
survey \citep{CO06.1}. The black dashed line indicates the shallowest
Schechter function consistent with the data of these two surveys
(SM). The blue solid line indicates the cumulative number counts
predicted by the semi-analytic model presented in \cite{SW08.1} for
SMGs with $S_{850~\mu\mathrm{m}} > 5$~mJy. Model predictions for SMGs
with $S_{850~\mu m} > 5$~mJy and $S_{\rm 1.4~GHz} > 30~\mu$Jy, and
SMGs with $S_{850~\mu m} > 5$~mJy and $S_{\rm 1.4~GHz} > 0.5~\mu$Jy
are indicated by the cyan solid and dashed lines, respectively. Note
that the cyan dashed line and the blue solid line are almost
indistinguishable.}
\label{fig:counts_swinbank}
\end{figure}

\begin{table}[t!]
\centering
\caption{Parameters for the $850~\mu$m differential number counts.}
\begin{tabular}{cccccc}
\hline\hline
\noalign{\smallskip}
Name          & $S_{*}$ (mJy) & $n_{0,*}^{(4)}$       & $\alpha$               & $\beta$  \\
\noalign{\smallskip}
\hline 
\noalign{\smallskip}
DB$^{(1)}$    & $9.6^{+0.3}_{-2.12}$ & $658\pm48$    & $2.12^{+0.14}_{-0.08}$ & $6.22^{+0.51}_{-0.34}$ \\
SB$^{(2)}$    & $4.30\pm0.08$        & $1039\pm69$   & $-2.62\pm0.10$         &  $-$                   \\
SM$^{(3)}$    & 4.22                 & 970           & -2.52                  &  $-$                   \\
\hline
\noalign{\smallskip}
\multicolumn{5}{l}{$^{(1)}$ best fit double power-law function, Eq. (\ref{eqn:dpl}).}\\
\multicolumn{5}{l}{$^{(2)}$ best fit Schechter function, Eq. (\ref{eqn:schechter}).}\\
\multicolumn{5}{l}{$^{(3)}$ shallowest Schechter function consistent with the data.}\\
\multicolumn{5}{l}{$^{(4)}$ expressed in deg$^{-2}$ for DB and in deg$^{-2}$ mJy$^{-1}$ for SB and SM.}\\ 
\end{tabular}
\label{submm_counts}
\end{table}
\noindent
Moreover, when fitting the observed cumulative number counts, they
added the supplementary constraint that the integrated light well
below $0.1$ mJy should not be higher than the extragalactic background
light \citep{PU96.1,FI98.1}. The resulting best-fit parameters are
summarized in Table \ref{submm_counts}.

\begin{figure*}[th!]
  \includegraphics[width=0.5\hsize]{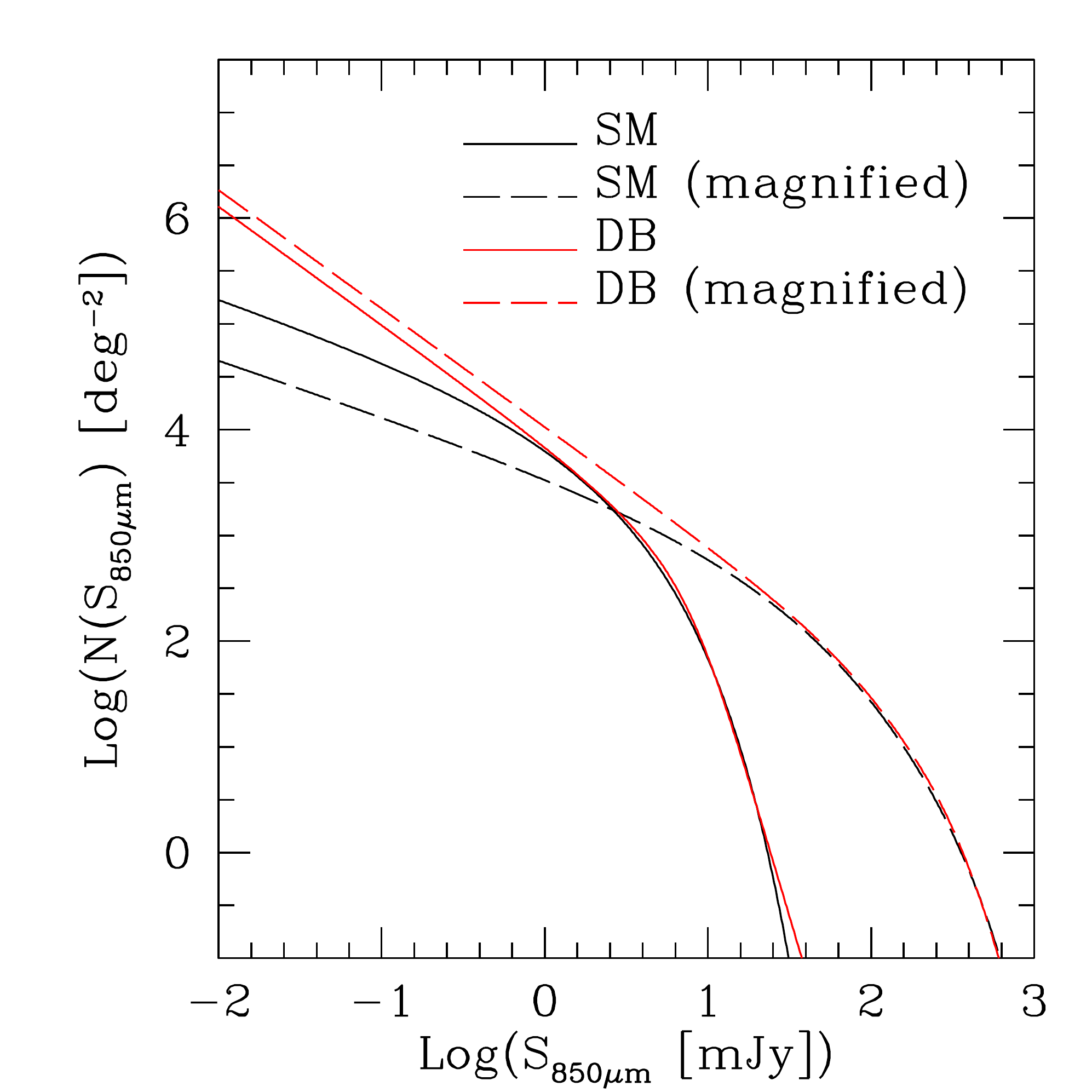}\hfill
  \includegraphics[width=0.5\hsize]{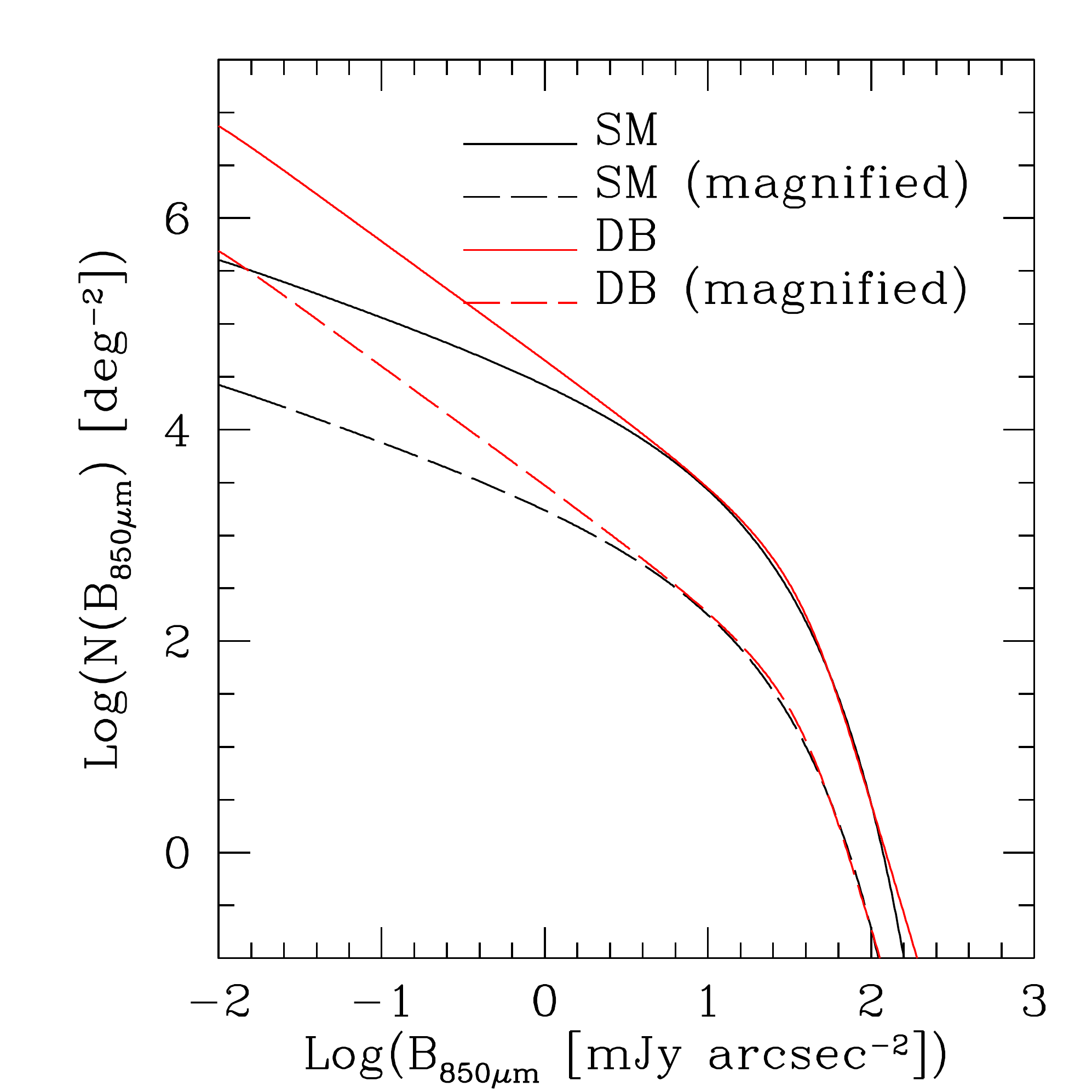}\hfill
  \includegraphics[width=0.5\hsize]{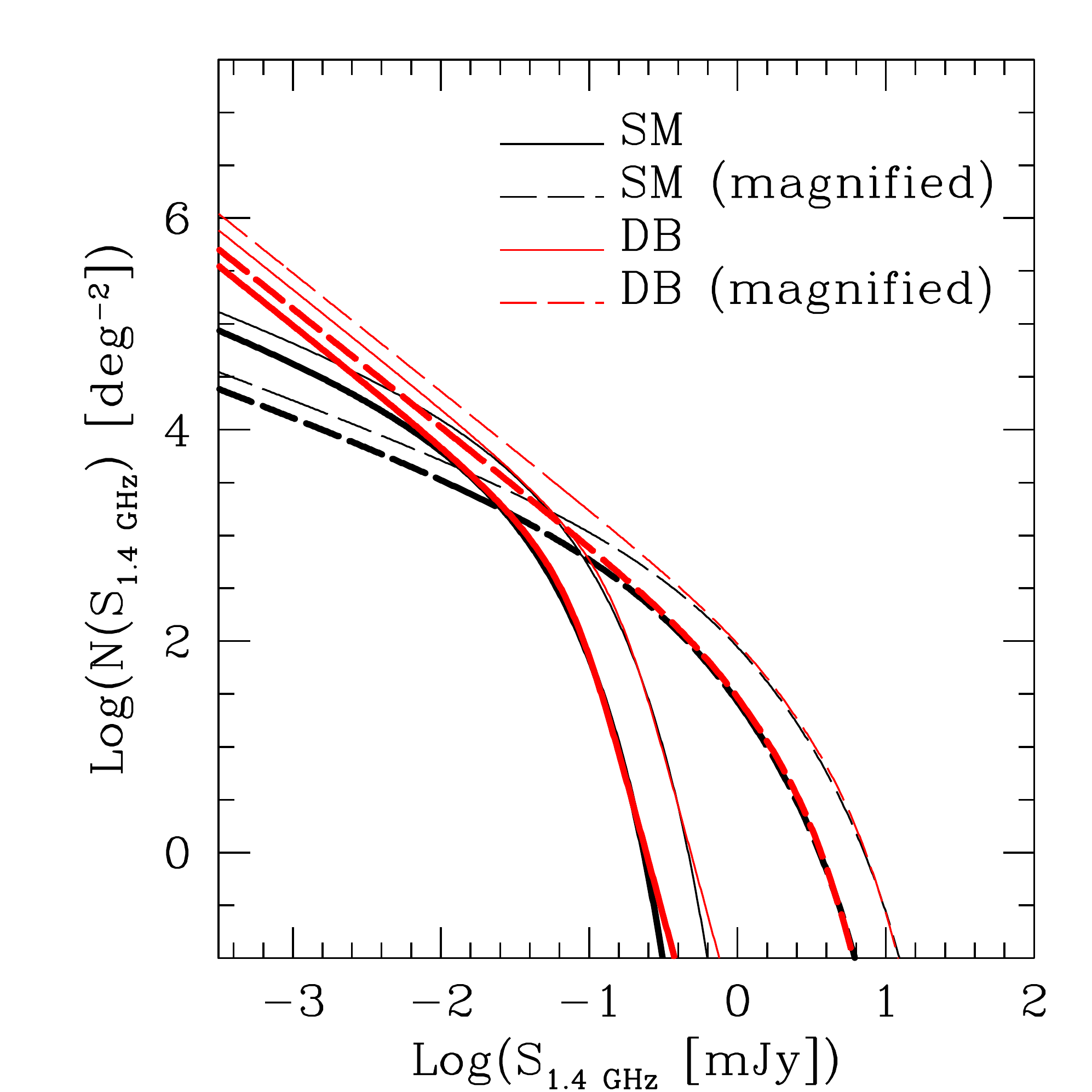}\hfill
  \includegraphics[width=0.5\hsize]{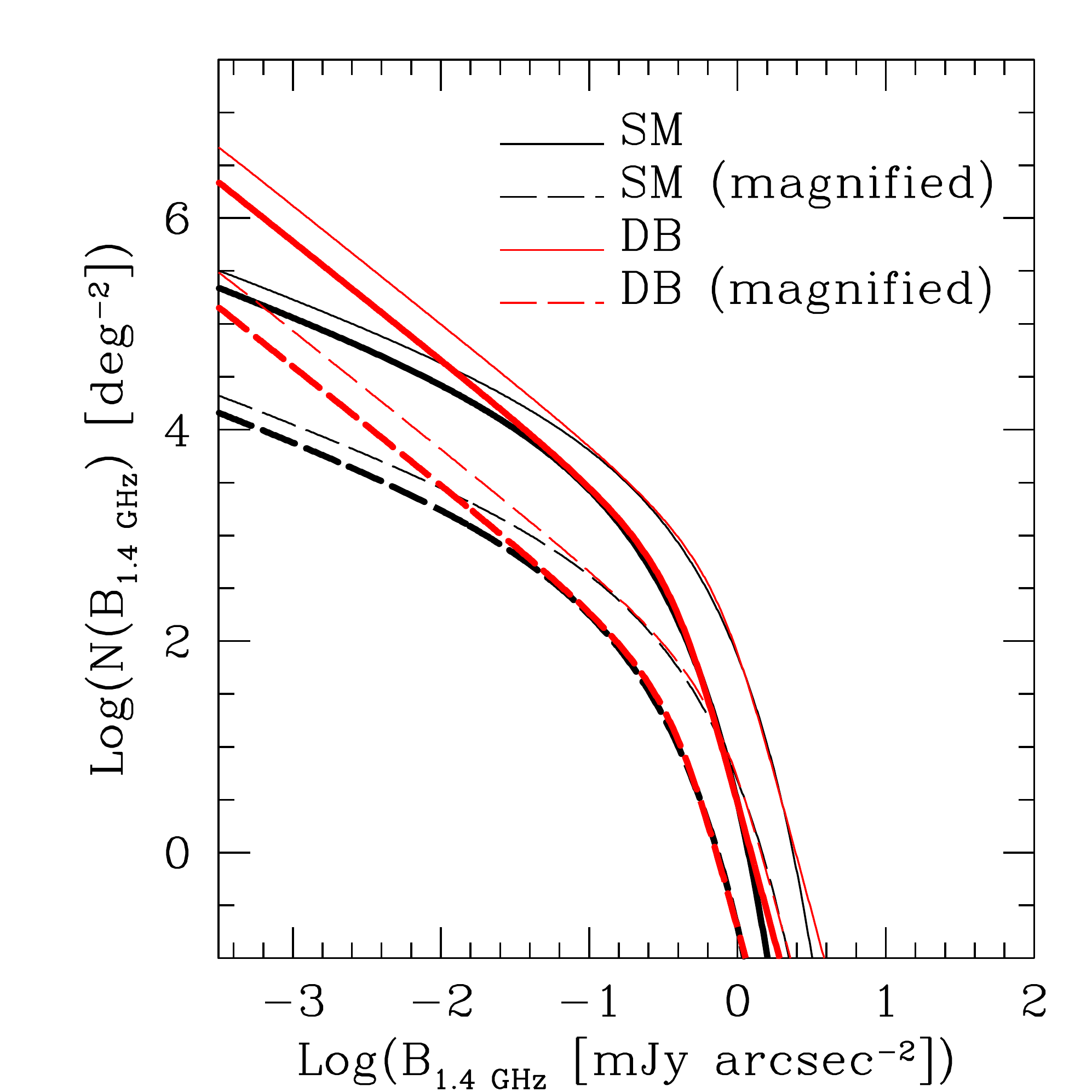}
\caption{The cumulative number counts of SMGs at $850~\mu$m (top
panels) and 1.4~GHz (bottom panels), as a function of flux density (left panels)
and surface brightness (right panels).  The conversion from flux density into
surface brightness was done assuming that the emission at both submm
and radio wavelengths is homogeneous and have $R_\mathrm{e} =
0.25\arcsec$. Thin lines correspond to $S_{1.4 \mathrm{GHz}} =
S_{850~\mu\mathrm{m}}/50$, while thick lines assume $S_{1.4
\mathrm{GHz}} = S_{850~\mu\mathrm{m}}/100$. The magnification pattern
used to compute the magnified number counts is the one for arcs with
$d \ge d_0 = 10$. SM indicates the shallowest Schechter function that is consistent with the data, while DB represents the best fit double power-law (see the text for more details).}
\label{fig:cc}
\end{figure*}

Figure \ref{fig:counts_swinbank} shows the (unmagnified)
cumulative number counts derived from the best fit Schechter function
(black solid line) and the best fit double power law function (red
solid line) presented in KN08 using Eq. (\ref{eqn:cum}). Note that,
whereas both curves behave almost identically at the bright flux end,
their predictions for the low flux number counts differ by a factor of
$\sim 2.5$ at $0.1$~mJy. Since the low flux tail of the submm counts
dominates the number of SMGs that could potentially be lensed, we
computed predictions for arcs produced by SMGs at $850~\mu$m for the
two following cases: (i) the shallowest Schechter function consistent
with the combined LSLS and SHADES data (also shown in Figure
\ref{fig:counts_swinbank}) and (ii) the best fit double power law
function, hereafter refered to as SM and DB,
respectively. The first one provides the minimum expected number of
arcs consistent with observations, whereas the second one gives the
number of arcs predicted by the best fit to the data (see Table
\ref{submm_counts}).

The cumulative number counts derived for these two cases as a function
of flux density are shown in the top left panel of Figure \ref{fig:cc}, including
the corresponding counts corrected for magnification bias using
Eq. (\ref{eqn:mag}). In the same Figure, the top right panel shows the
cumulative number counts as function of surface brightness. The
corresponding counts corrected for magnification bias (which will be
used to compute the number of arcs) were derived using
Eq. (\ref{eqn:mag_B}).

\subsubsection{Radio number counts}

Figure \ref{fig:counts_swinbank} also shows the cumulative submm
number counts predicted by the SW08 model (blue solid line)
compared with the results from different $\rm 850~\mu$m SCUBA
surveys. Note that, although the model tends to over-predict the
counts at faint fluxes compared with the best fits provided by KN08
(red and black solid lines), it is consistent with the observational
errors. The cyan solid line indicates the predicted counts for SMGs
with radio counterparts assuming $S_{\rm 1.4~GHz} > 30~\mu$Jy. The
fact that its shape is different from the shape of the blue solid
curve is because current observations only detect radio emission from
$\sim 60 \%$ of the observed SMGs. However, if we allow the
sensitivity threshold to go down to the $\mu$Jy level expected for
\emph{e}-MERLIN, the SW08 model indicates that it would be
possible to detect all the radio counterparts of SMGs with
$S_{850~\mu\mathrm{m}} > 5$~mJy (cyan dashed line).

Since SMGs seem to follow the FIR/radio correlation
\citep[e.g.][]{KO06.1}, the $\rm 1.4~GHz$ number counts of SMGs (which
we need to predict the number of radio arcs) could be derived by
scaling the $850~\mu$m number counts introduced in the previous
section (DB and SM). As shown in Figure 7 of CH05, the ratio between
the $850~\mu$m flux density and the $1.4$~GHz flux density shows a
broad scatter (up to one order of magnitude), which is probably a
consequence of the strong influence of the dust temperature on the
SEDs of SMGs. However, most of the points in this Figure are located
between redshift 2 and 3, and have an average $S_{850~\mu\mathrm{m}} /
S_{1.4~\mathrm{GHz}}$ ratio between $50$ and $100$. Therefore, we
decided to use these two scaling factors to derive first order upper
and lower limits of the radio number counts of SMGs. The resultant
cumulative radio number counts are shown in the lower panels of Figure
\ref{fig:cc}. Note that the values of $50$ and $100$ chosen for
the submm/radio flux density ratio are meant to be indicative, since
there are still many sources that display a ratio below $50$ or above
$100$. The reader interested in results given by different values of
this ratio can scale the curves appropriately in the upper panels of
Figure \ref{fig:cc}. Also, exact numerical values can be made
available by the authors upon request.

\section{Strong lensing optical depth}\label{sct:model}

To compute the total optical depth for lensed SMGs, we constructed a
synthetic cluster population composed of $q = 1000$ cluster-sized
dark-matter halos with masses uniformly distributed in the interval
$[10^{14},2.5 \times 10^{15}]~M_\odot h^{-1}$ at $z=0$. Note that it
is not necessary to extract these masses according to the cluster mass
function, since this is already taken into account in
Eq. (\ref{eqn:tau}) by weighting the cross sections with the function
$n(M,z)$. The structure of each cluster is modeled using the NFW
density profile (\citealt{NA95.1,NA96.1,NA97.1}), which constitutes a
good representation of average dark-matter halos over a wide range of
masses, redshifts and cosmologies in numerical $n$-body simulations
\citep{DO04.1}. Several studies of strong lensing and X-ray luminous
clusters also show that these are well fitted by an NFW profile
\citep{SC07.2,OG09.1}. This profile also has the advantage that its
lensing properties can be described analytically \citep{BA96.1}.

To account for the asymmetries of real galaxy clusters, the halos are
assumed to have elliptically distorted lensing potentials. However,
instead of considering a single ellipticity value to describe all the
synthetic cluster lenses, we derived an ellipticity distribution from
a set of numerical clusters extracted from the \emph{MareNostrum}
simulation \citep{GO07.1}. The strong lensing analysis required to
generate this ellipticity distribution was taken from Fedeli et al.
(2009, in preparation), as described in Section \ref{sct:shape}. For
each simulated cluster, the lensing analysis along three orthogonal
projections was performed, computing the cross sections for arcs with
$d \ge d_0 = 7.5$ and sources with $z_\mathrm{s} = 2$. For each of
these projections, we found the ellipticity $e$ of the NFW lens whose
cross section is closest to the cross section of the numerical
cluster, i.e., we found the ellipticity that minimizes the quantity

\begin{equation}\label{eqn:dist}
r(e) = \left|\sigma_{7.5}^{\mathrm{(n)}} - \sigma_{7.5}(e)\right|, 
\end{equation}
where $\sigma_{7.5}^\mathrm{(n)}$ is the cross section of the numerical
lens, and $\sigma_{7.5}(e)$ is that of the NFW lens for a given potential ellipticity $e$.

\begin{figure}[t]
  \includegraphics[width=\hsize]{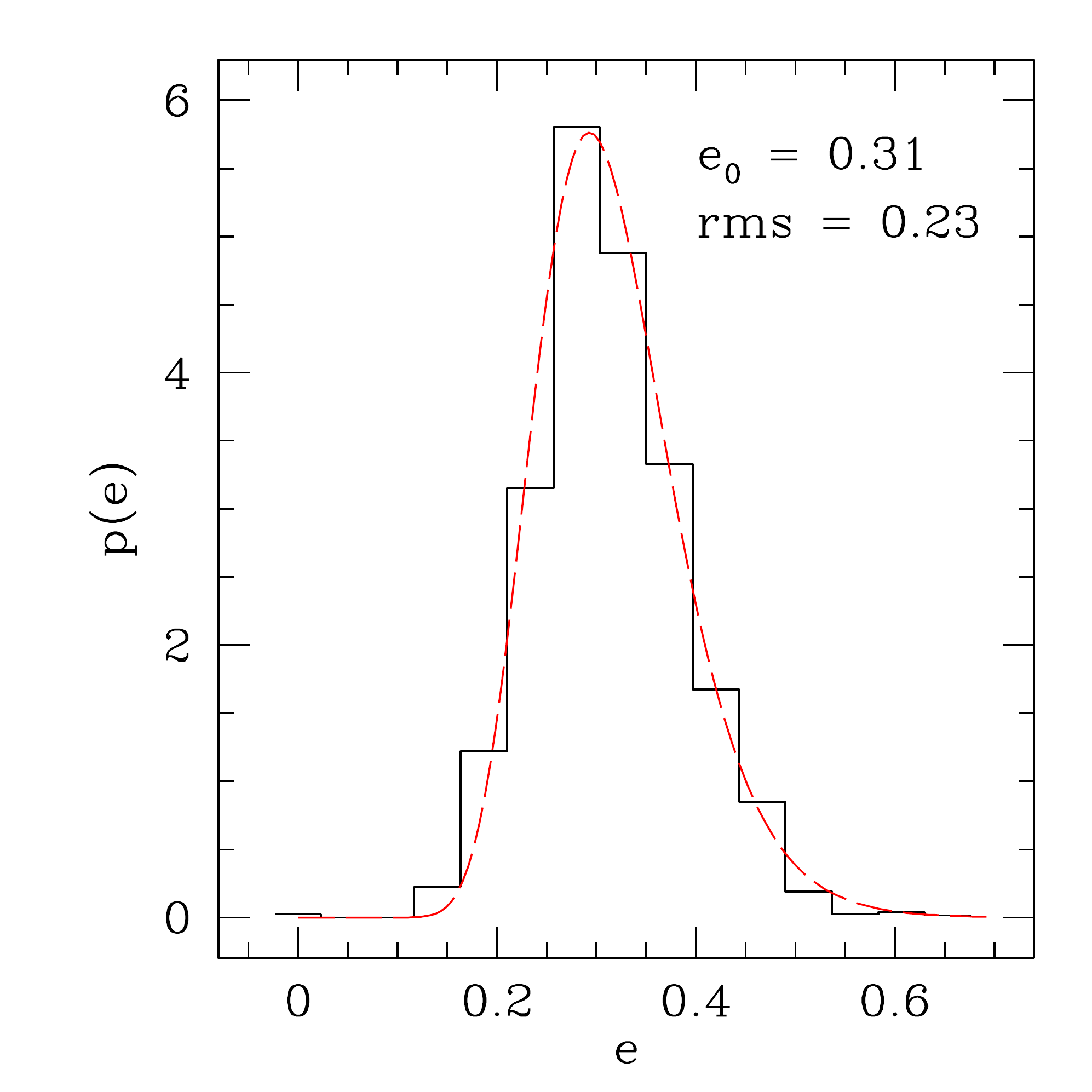}\hfill
\caption{The distribution of NFW lens ellipticities fitting the cross sections of
a sample of numerical clusters. The red dashed line
represents the best fit log-normal distribution, whose median and
dispersion are labeled in the top-right corner of the plot.}
\label{fig:ed}
\end{figure}

Figure \ref{fig:ed} shows the distribution of the ellipticities that
minimize the quantity $r(e)$ in Eq. (\ref{eqn:dist}) for all clusters
in our numerical analysis. The dashed red line is derived by fitting
the distribution with a log-normal function of the kind

\begin{equation}\label{eqn:lognormal}
p(e)de = \frac{1}{\sqrt{2\pi}\sigma_\mathrm{e}} \exp\left[ -\frac{\left[\ln(e) - \ln(e_0)\right]^2}{2\sigma_\mathrm{e}^2}\right] d\ln(e),
\end{equation}
where the best-fit parameters are $e_0 = 0.31$ and $\sigma_\mathrm{e}
= 0.23$. The ellipticity values used to
characterize the potential of the synthetic NFW cluster lenses were then
randomly extracted from the above distribution.

Elliptical NFW profiles are a good representation
of realistic cluster lenses only when the clusters do not undergo
major merger events \citep{ME03.1}. Since the merger
activity of galaxy clusters is known to have a significant effect on
the statistics of giant arcs \citep{TO04.1,FE06.1}, it has to be taken
into account in the construction of the synthetic cluster
population. For this reason, we used the excursion set
formalism\footnote{Also referred to as the "extended \cite{PR74.1}
formalism"} developed by \cite{LA93.1} (see also
\citealt{BO91.1,SO99.1}) to construct a backward merger tree for each
model cluster at $z = 0$, assuming that each merger is binary
(see discussion in \citealt{FE07.1}). When a merger happens, the event
is modeled assuming that the two merging halos (also described as
elliptical NFW density profiles) approach each other at a constant
speed. The duration of the merger is set by the dynamical timescales
of the two halos (see \cite{FE07.1} and \cite{FE08.1}
for a detailed description of the modeling procedure).

With the synthetic cluster population constructed in this way, the
total average optical depth was derived by computing individual cross sections with the semi-analytic
algorithm developed by \cite{FE06.1}, especially designed to estimate
the strong lensing cross sections of individual lenses in a fast and
reliable way. The optical depth for a discrete set of lenses can be recast as

\begin{equation}\label{eqn:discrete}
\tau_{d_0}(z_\mathrm{s}) = \frac{1}{4\pi D_\mathrm{s}^2} \int_0^{z_\mathrm{s}} \left[ \sum_{i=1}^{q-1} \bar{\sigma}_{d_0,i}(z) \int_{M_i}^{M_{i+1}} n(M,z) dM \right] dz ,
\end{equation}
where the masses $M_i$ ($1 \le i \ne q$) have to be sorted from the
lowest to the highest at each redshift step, and the quantity
$\bar{\sigma}_{d_0,i}(z)$ is defined as

\begin{equation}
\bar{\sigma}_{d_0,i}(z) \equiv \frac{1}{2} \left[ \sigma_{d_0}(M_i,z) + \sigma_{d_0}(M_{i+1},z) \right].
\end{equation}
This effectively means that, for all the clusters with mass
between $M_i$ and $M_{i+1}$, we assume the average cross section of
the model dark-matter halos with masses $M_i$ and $M_{i+1}$. The
algorithm of \cite{FE06.1} for computing strong lensing cross sections
consists of first assuming sources as point-like circles, and then
introducing the effect of source ellipticity according to
\cite{KE01.1}. The source finite size is taken into account by
convolving the local lensing properties over the typical source size.

The total average optical depth is calculated by integrating
Eq. (\ref{eqn:discrete}) over the source redshift
distribution. Effectively, the $p(z_\mathrm{s})$ weighting is avoided
since we assigned individual source redshifts (randomly extracted from
the distribution $p(z_\mathrm{s})$) at each of the $q$ halos in the
cluster sample and evolved their merger trees back in time until the
respective source redshift. Given the large number of synthetic
dark-matter halos used, this approach allows one to omit $p(z_\mathrm{s})$
in Eq. (\ref{eqn:av}) when the redshift integral is discretized.

Despite the fact that the ellipticity distribution used to model the
synthetic cluster population was derived for $d \ge d_0 = 7.5$, the
best fit parameters of Eq. (\ref{eqn:lognormal}) can be used to
compute the cross sections for arcs also with $d_0 = 10$ without
compromising the results. The reason is that, although there might be
a mild dependence of the distribution of lensing ellipticities on
$d_0$, it is the overall caustic structure that defines the abundance
of arcs above a certain $d_0$, regardless of its precise value. In
fact, the criterion used to determine the ellipticity distribution is
based on the similarity between the cross sections of the NFW lens and
the numerical lens, which is an indirect way of comparing the overall
caustic structures produced by both kinds of lenses. To verify this
argument, the ellipticity distribution was re-computed using a
criterion that is directly related to the caustic structure, that is,
by defining the best fitting ellipticity as the one that minimizes the
modified Hausdorff distance\footnote{This parameter constitutes one of
the best ways to quantify the morphological difference between two
sets of points \citep{DU94.1}. See also \citealt{RZ07.1} for a
different application to gravitational lensing.} between the critical
lines of the numerical and NFW lenses. The ellipticity distribution
obtained in this way is very similar to the one depicted in Figure
\ref{fig:ed} ($e_0 = 0.30$, $rms = 0.34$).

The median of the ellipticity distribution derived in this work using
the lensing cross section ($e_0 = 0.31$, see Figure \ref{fig:ed}) is
fully consistent with the one obtained by \cite{ME03.1} comparing the
deflection angle maps ($e \sim 0.3$, using lenses placed at $z \sim
0.3$ and a source population at $z_\mathrm{s} = 1$). Even though these
two criteria are arguably tightly related, the former is a quantity
that is more directly related to observables than the latter, hence it
is reassuring that they give comparable results. In addition, this
result extends the previous one by quantifying the scatter around the
median ellipticity and dealing with lenses that are distributed in
redshift.

\section{Results} \label{sct:results}

\subsection{Arc redshift distribution }

\begin{figure*}[t]
  \includegraphics[width=0.5\hsize]{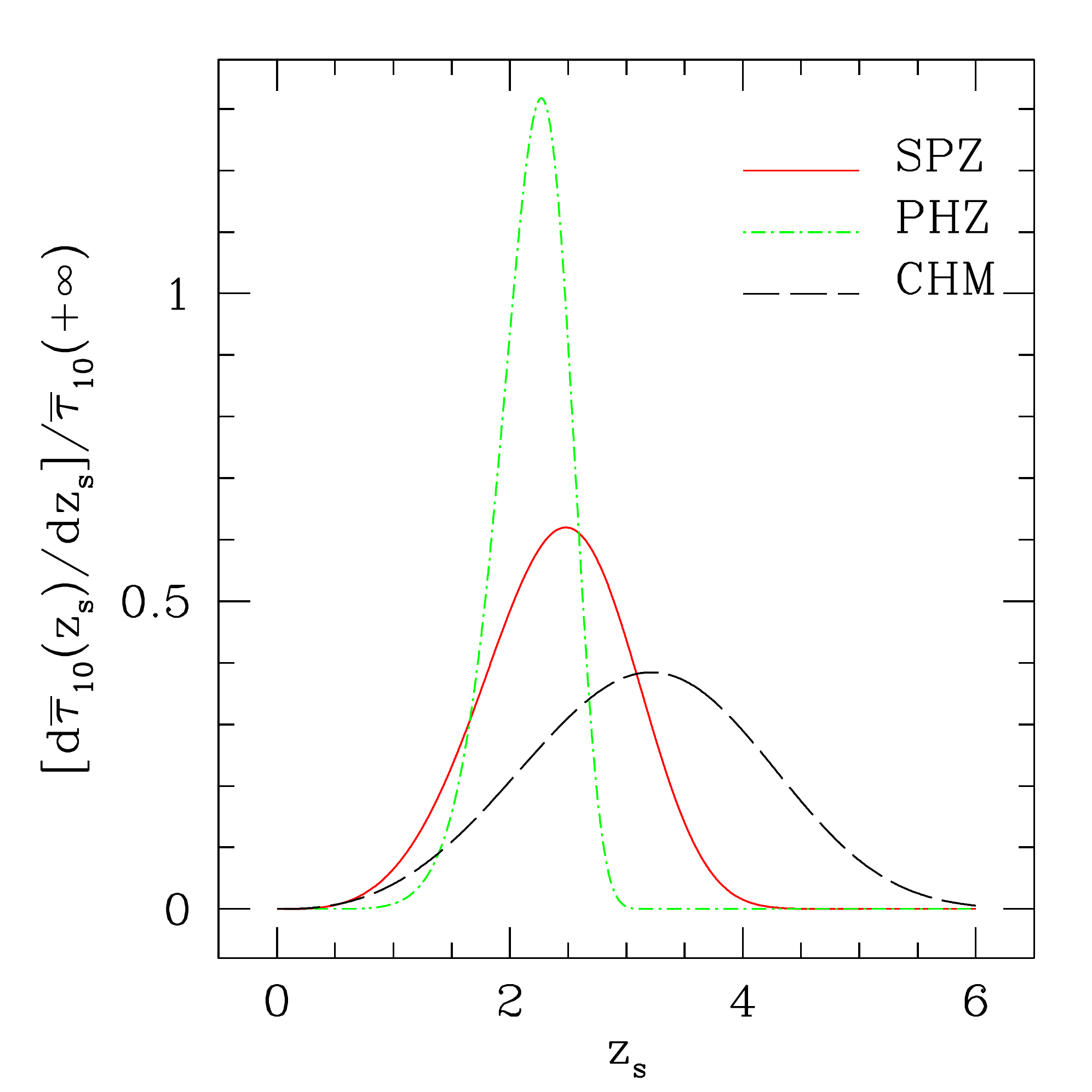}
  \includegraphics[width=0.5\hsize]{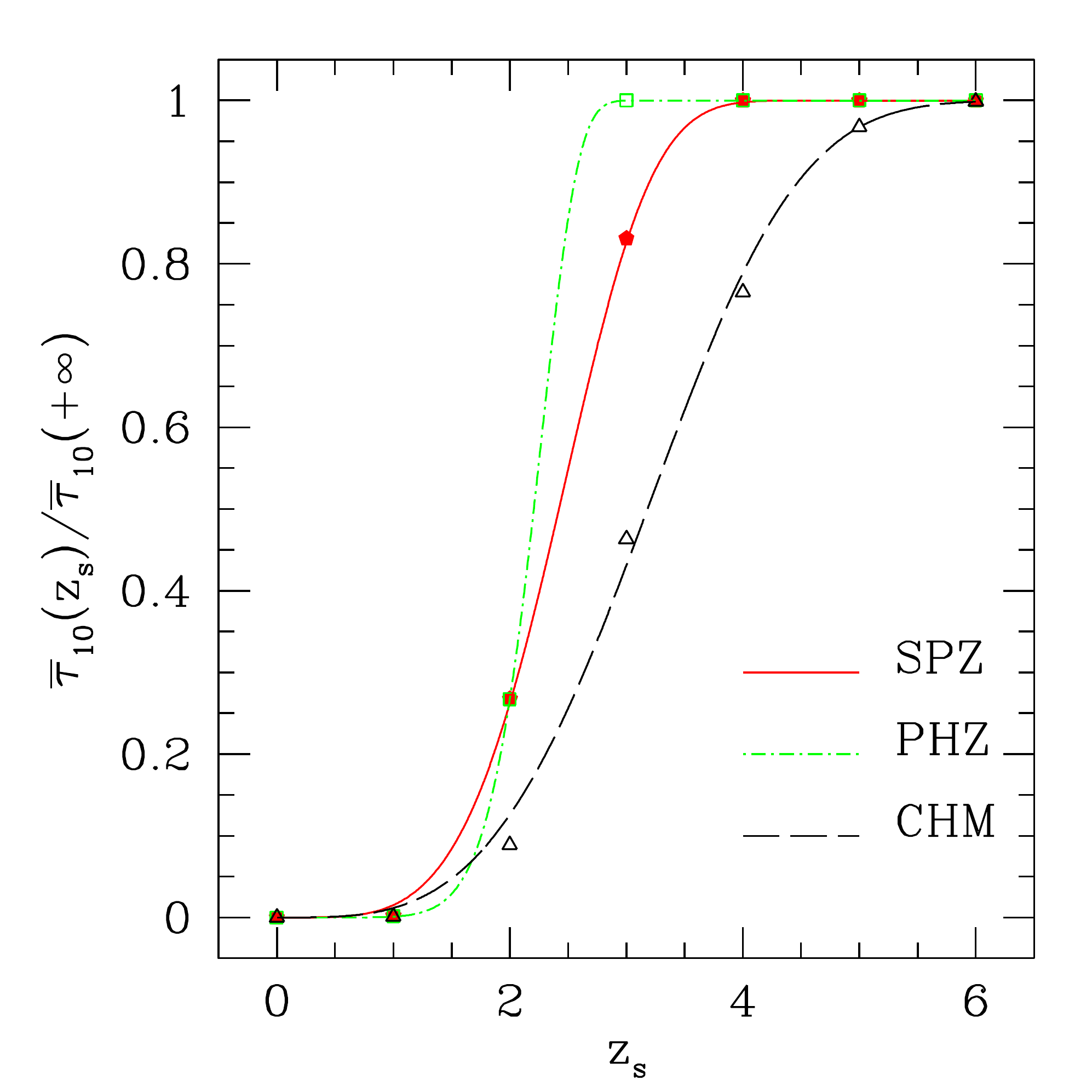}\hfill
\caption{\emph{Left panel}. Differential redshift distributions of
SMGs producing large gravitational arcs, corresponding to the
cumulative distributions reported in the right panel.  \emph{Right
panel}. The cumulative redshift distributions of SMGs producing radio
arcs with $d \ge d_0 = 10$ for input source redshift distributions SPZ
, PHZ and CHM (see the text for more details). Lines are the best fit
functions given in Eq. (\ref{eqn:zArc}) and Table \ref{tab:zArc},
according to the labels in the plot.}
\label{fig:op}
\end{figure*}

The redshift distribution of arcs with $d \ge d_0$ (arc redshift
distribution hereafter) is expected to provide information about the
redshift distribution of the background source population that is
being lensed. However, it will also be distorted by the fact that the
abundance of massive and compact galaxy clusters evolves with
redshift, and that the lensing efficiency depends on the relative
distances of sources and lens with respect to the observer.

To assess the potential of this approach to gather
information about the intrinsic redshift distribution of SMGs, we
derived the arc redshift distribution associated with each of the
three source redshift distributions used as inputs in our calculations
(see Figure \ref{fig:zd}). This was done by computing, for each input
distribution, the average optical depth
$\bar{\tau}_{d_0}(z_\mathrm{s})$ for several different values of
$z_\mathrm{s}$. That is, we excluded in the computation of the average
optical depth those model clusters (and their respective chain of
progenitors) whose associated source redshifts were $>
z_\mathrm{s}$. The resultant cumulative arc redshift distributions,
obtained after normalizing the optical depth for each $z_\mathrm{s}$
to the total average optical depth $\bar{\tau}_{d_0}(+\infty) =
\bar{\tau}_{d_0}$, are shown in the right panel of Figure
\ref{fig:op}. The lines correspond to the best fit for each
distribution provided by the simple function

\begin{equation}\label{eqn:zArc}
\frac{\bar{\tau}_{d_0}(z_\mathrm{s})}{\bar{\tau}_{d_0}(+\infty)} = 1 - \exp \left( -\frac{z_\mathrm{s}}{z_{\mathrm{s},*}}\right)^\gamma,
\end{equation}
where $z_{\mathrm{s},*}$ indicates where the transition between the extrema $0$ and $1$ occur,
and $\gamma$ indicates how sharp this transition is. The best-fit
parameters for each of the three input source redshift distributions
are summarized in Table \ref{tab:zArc}. The corresponding differential
arc redshift distributions are shown in the left panel of
Figure \ref{fig:op}.

\begin{table}[t!]
  \caption{Parameters of the arc redshift distributions shown in Figure \ref{fig:op}.}
  \label{tab:zArc}
  \begin{center}
    \begin{tabular}{ccc}
      \hline\hline
      \noalign{\smallskip}
      $p(z_\mathrm{s})$ & $z_{\mathrm{s},*}$ & $\gamma$\\
      \noalign{\smallskip}
      \hline
      \noalign{\smallskip}
      SPZ & $2.64$ & $4.31$ \\
      PHZ & $2.31$ & $8.20$ \\
      CHM & $3.53$ & $3.53$ \\
      \hline
    \end{tabular}
  \end{center}
\end{table}

Once more, these arc redshift distributions have been shown for arcs
with length-to-width ratio larger than $d_0 = 10$ only. As we verified, since the relative contribution of individual model clusters to the total average optical depth is the same for both $d_0 = 7.5$
and $d_0 = 10$, the resulting arc redshift distribution also does not change significantly between the two choices.

\begin{figure*}[t]
  \includegraphics[width=0.5\hsize]{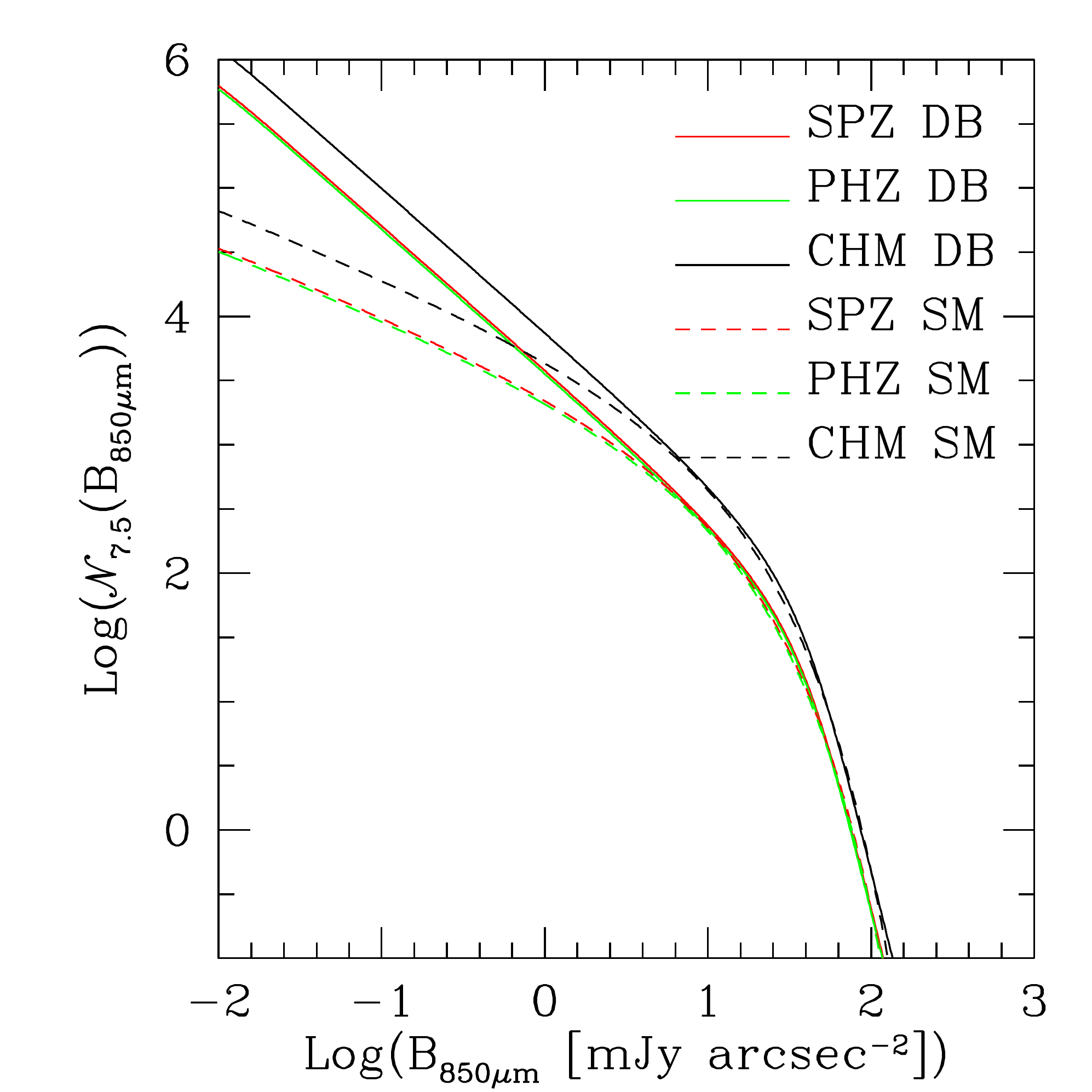}\hfill
  \includegraphics[width=0.5\hsize]{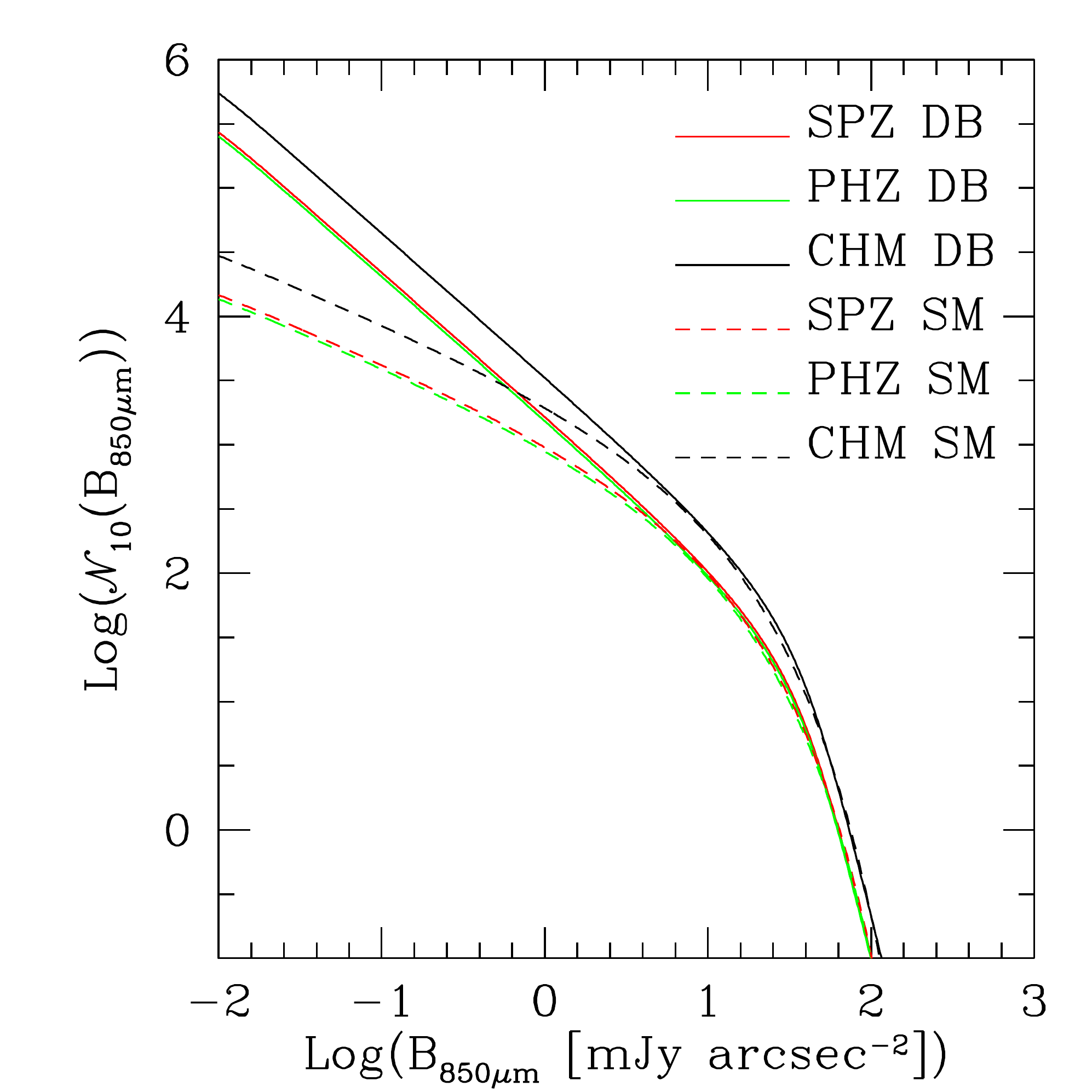}\hfill
  \includegraphics[width=0.5\hsize]{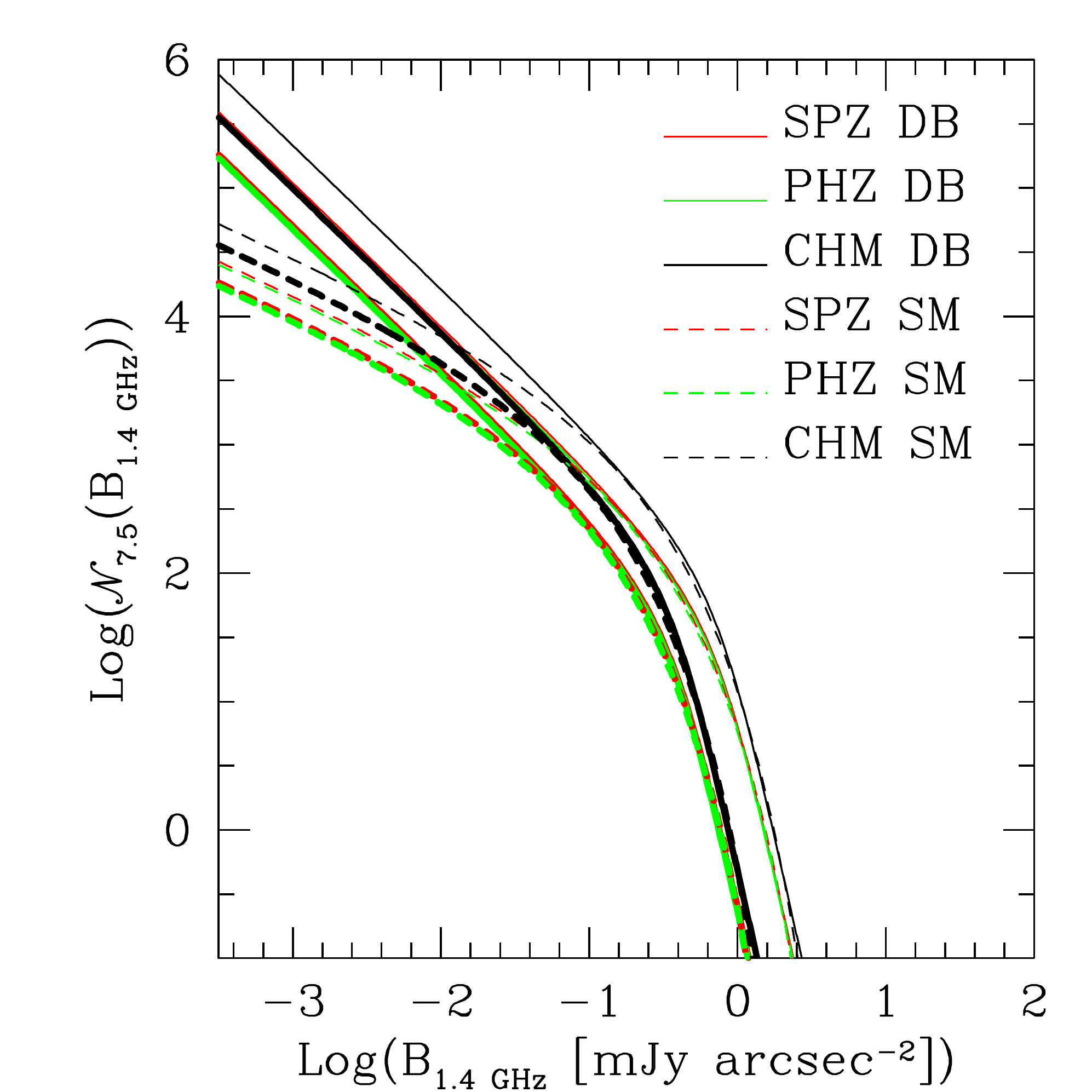}\hfill
  \includegraphics[width=0.5\hsize]{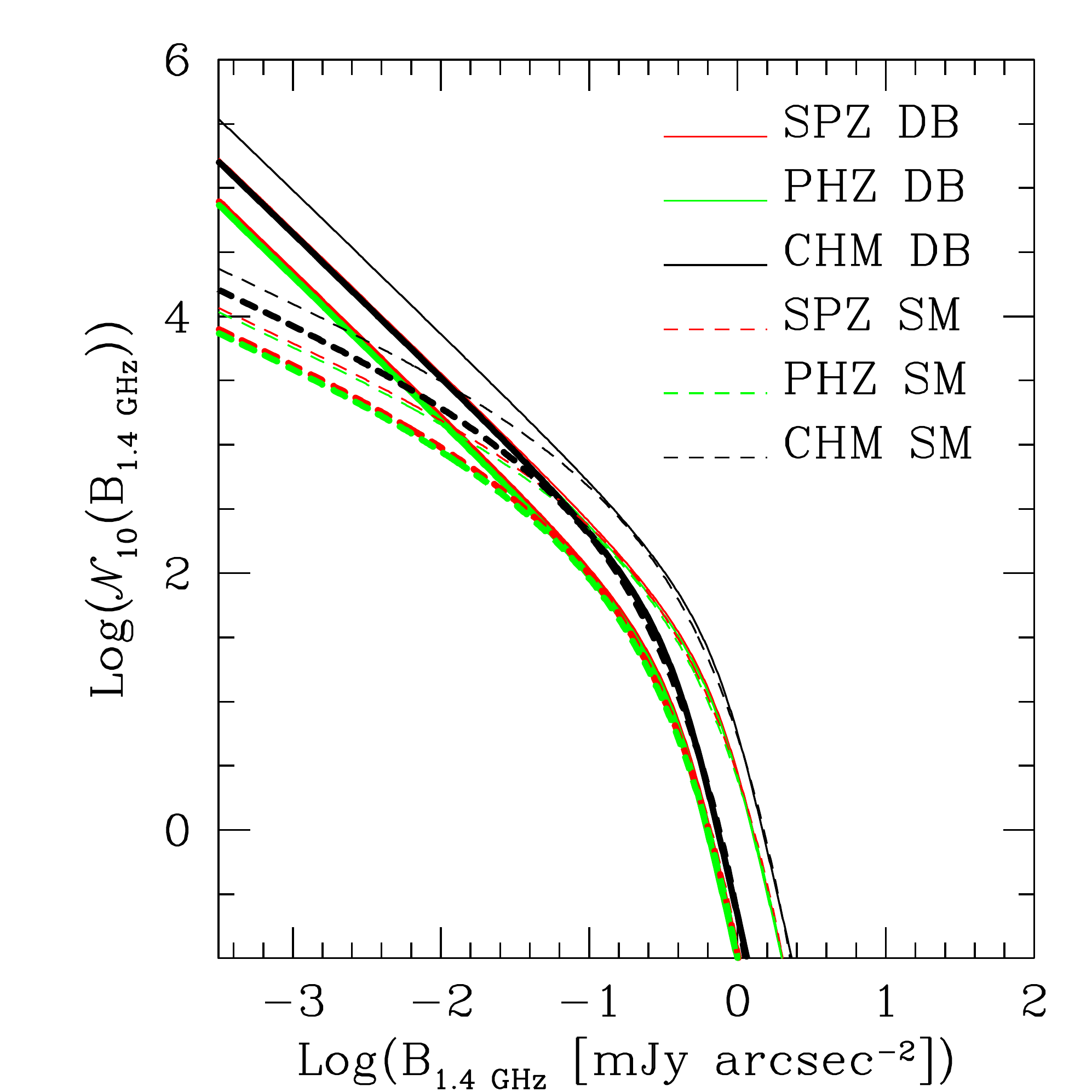}
\caption{The total number of arcs with $d \ge d_0 = 7.5$ (left panels)
and $d \ge d_0 = 10$ (right panels) that are predicted to be observed
in the whole sky above the surface brightness reported on the
abscissa. Results for each of the three input source redshift
distributions as well as both source number counts adopted in this work are shown, according to the
labels. The two top panels refer to surface brightness at $850~\mu$m,
while the bottom panels refer to $1.4$~GHz. In the bottom panels, thin lines refer to $S_{1.4~\mathrm{GHz}} = S_{850~\mu m}/50$, while thick lines refer to $S_{1.4~\mathrm{GHz}} = S_{850~\mu m}/100$. Please note the difference in scale on the abscissa between top and bottom panels.}
\label{fig:arcCounts}
\end{figure*}

\begin{figure*}[t]
  \includegraphics[width=0.5\hsize]{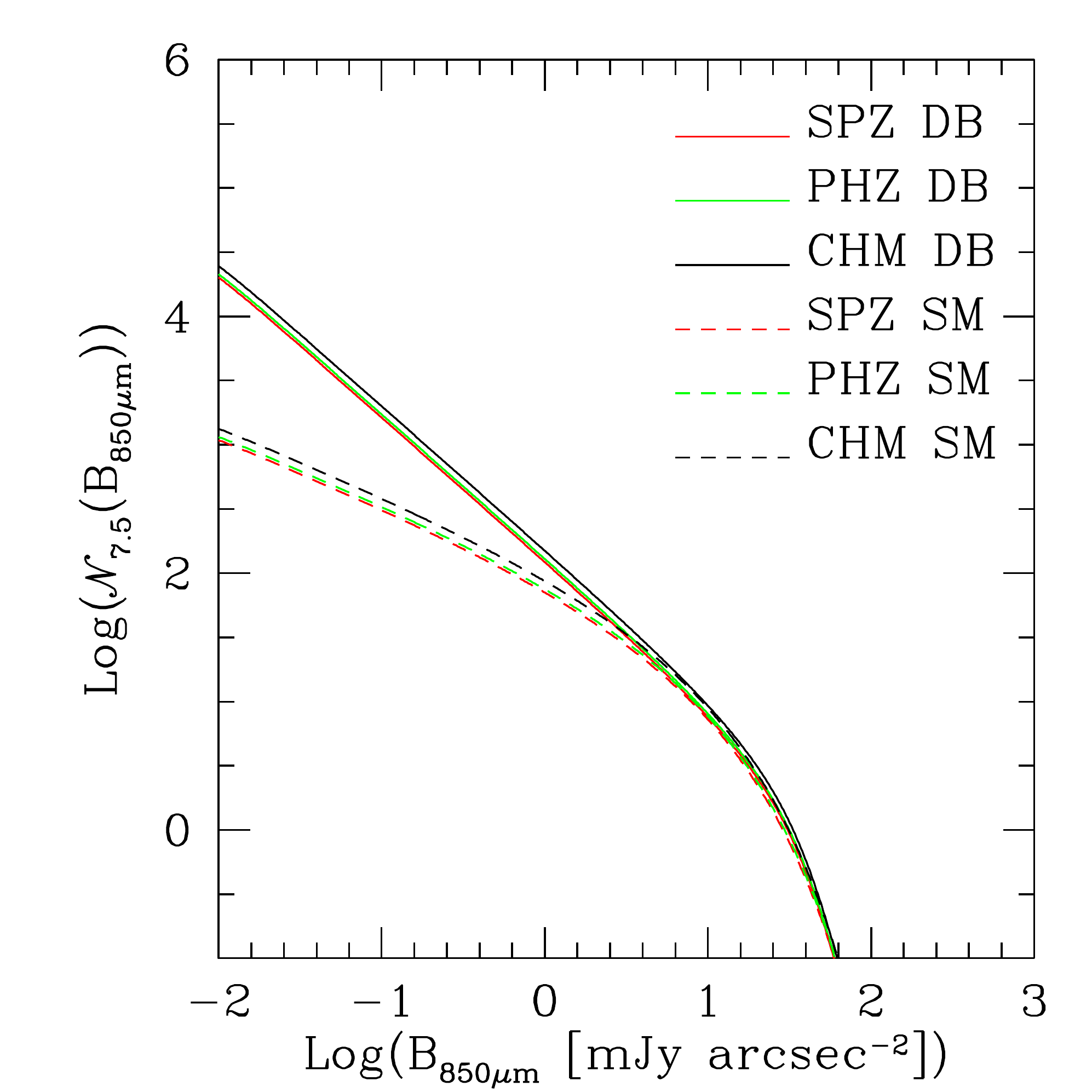}\hfill
  \includegraphics[width=0.5\hsize]{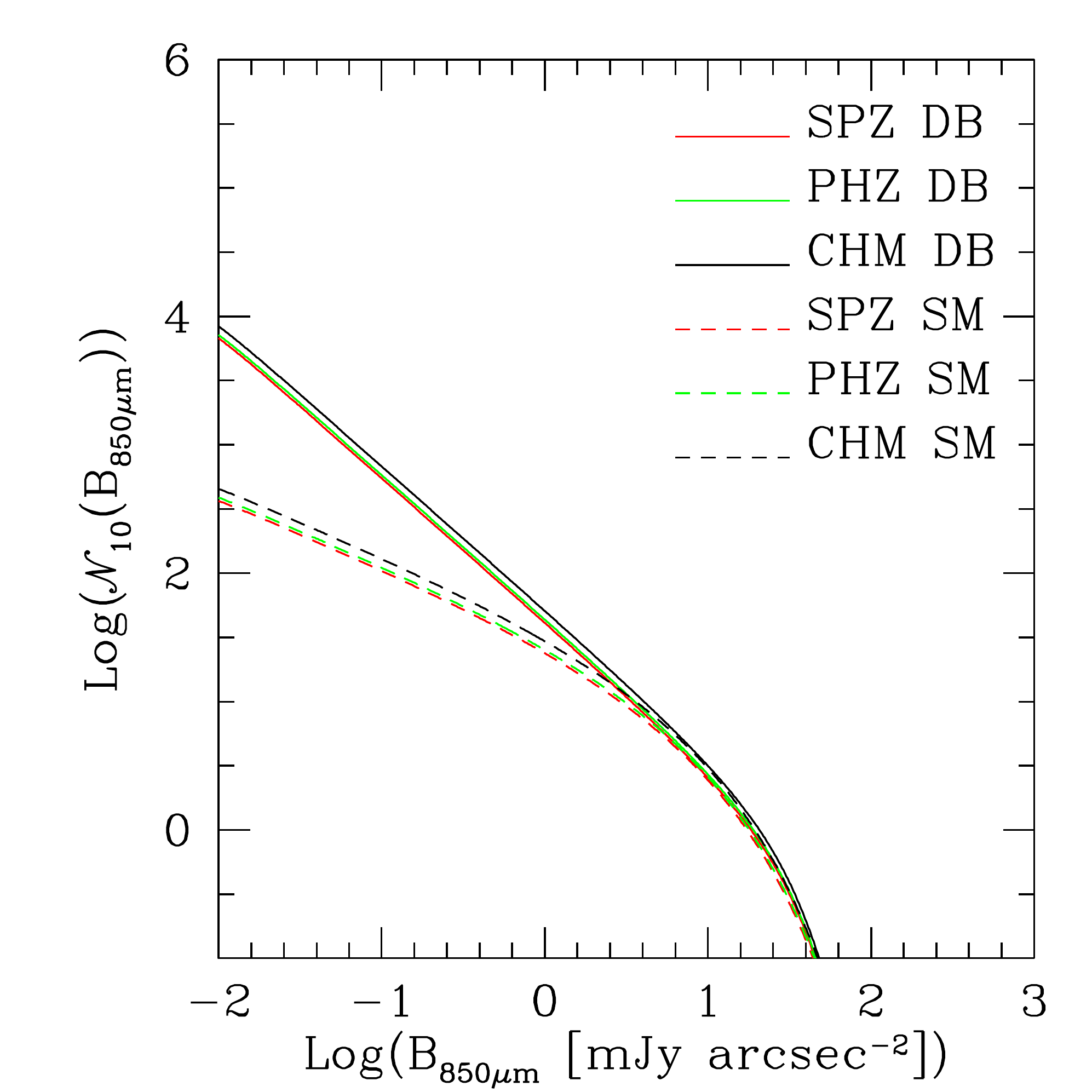}\hfill
  \includegraphics[width=0.5\hsize]{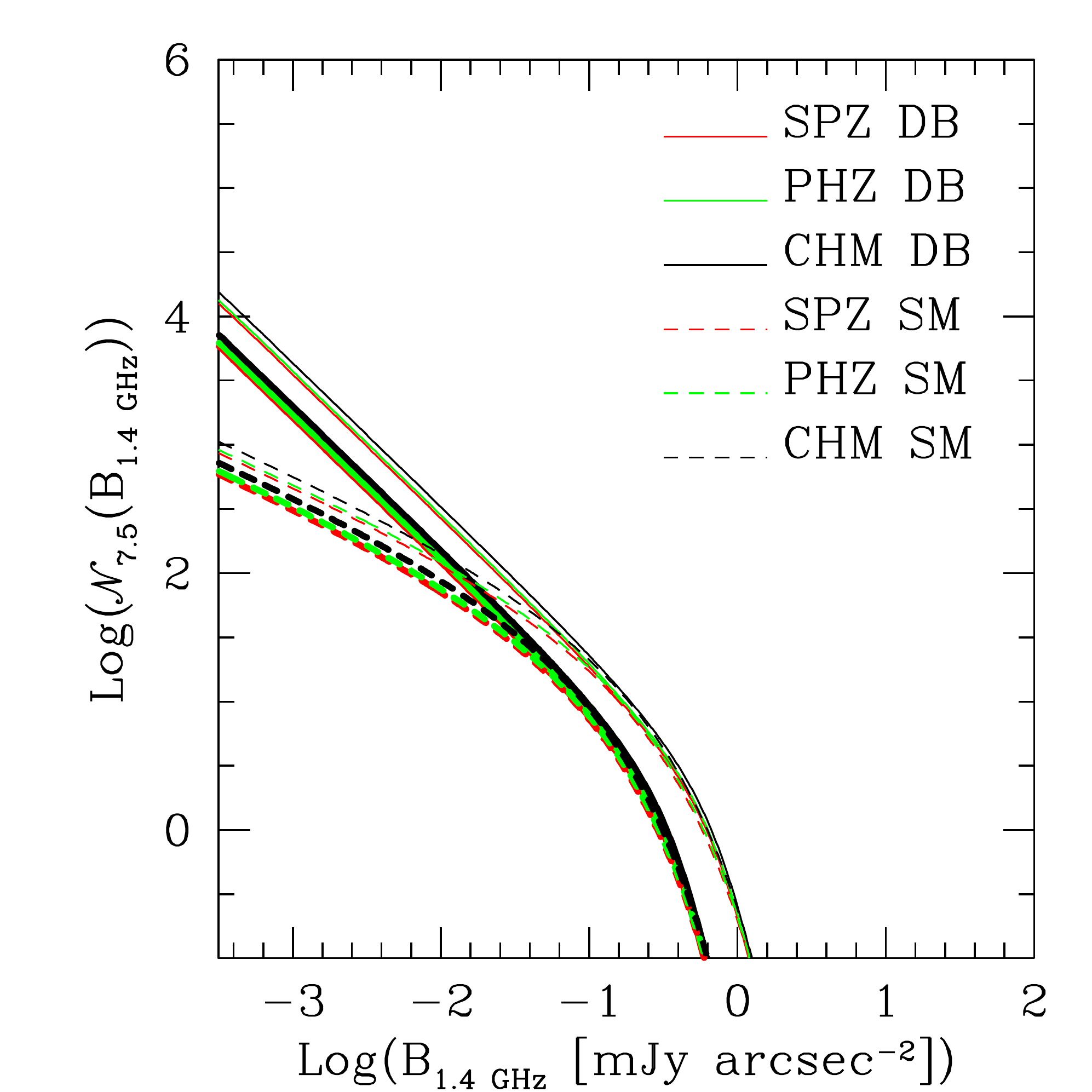}\hfill
  \includegraphics[width=0.5\hsize]{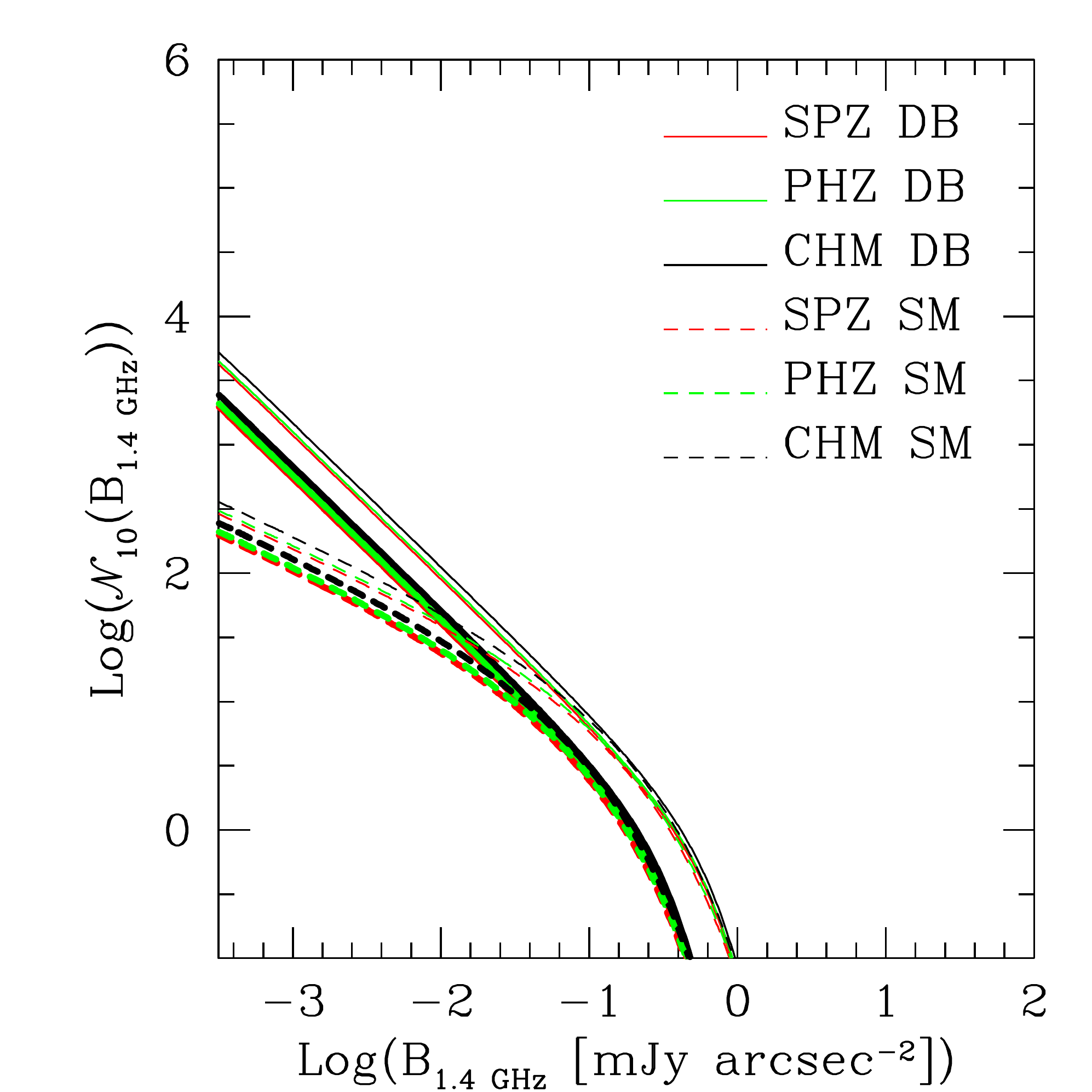}
\caption{The total number of arcs with $d \ge d_0 = 7.5$ (left panels)
and $d \ge d_0 = 10$ (right panels) that are predicted to be observed
in the whole sky above the surface brightness reported on the abscissa. Only
clusters with mass $M \ge 5 \times 10^{14} M_\odot h^{-1}$ are
included in the calculations. Results for each of the three input
source redshift distributions and both source number counts adopted in this work are shown,
according to the labels. Top panels refer to submillimeter number
counts, while bottom panels refer to radio number counts (notice the
different scale on the horizontal axis). The difference between thin and thick lines is as in Figure \ref{fig:arcCounts}.}
\label{fig:arcCounts_mass}
\end{figure*}

A comparison between Figure \ref{fig:op} and Figure \ref{fig:zd} shows
that, as expected, the arc redshift distributions reflect the general
properties of the source redshift distributions used as input, but
there are also some noteworthy differences between them. For instance,
the arc redshift distribution associated with CHM tends to zero at
very low redshift, unlike in the case of the original CHM distribution. The reason is that low
redshift sources do not produce many arcs, because (i) they have
very few potential lenses at their disposal, and (ii) the lensing efficiency
of those lenses is very low due to geometric suppression. This
results in a lack of low redshift arcs in the distribution, which
shifts its peak to higher redshifts compared with the CHM peak (from
$\rm z_{p}=2.3$ to $\rm z_{p} \sim 3.2$). Similarly, the peak of the arc
redshift distribution corresponding to SPZ is shifted from
$z_\mathrm{p} = 1.76$ to $z_\mathrm{p} \gtrsim 2$.

On the other hand, the arc redshift distribution corresponding to PHZ
is not significantly shifted but visibly narrowed.  As in the previous
cases, low-redshift sources are removed from the distribution because
they cannot produce arcs, but the distribution could not shift at
higher redshift because the input source redshift distribution is
immediately truncated at $z_\mathrm{s} \lesssim 3$. In other words,
there is too little room between the drop due to lensing
efficiency and the one due to the cutoff of the input distribution to
allow a significant shift in its peak, and the only possible
consequence for the distribution is to shrink and increase the peak
height in order to preserve the normalization.

In general, it is apparent that the differences between different
source redshift distributions are somewhat enhanced when it comes to
the arc redshift distribution. Therefore, assuming that redshift
information is available for arcs, this approach can in principle be
used to obtain some information about the general characteristics of
the source redshift distribution, although it will probably not allow one to
distinguish between redshift distributions that are very similar.

\subsection{Number of radio and submm arcs} \label{sect:arc_results}

In this section we present and discuss the main results of this work:
the predicted number of arcs produced by SMGs at radio and submm
wavelengths.  To that end, we computed the total average optical depth
for each of the three source redshift distributions presented in
Section \ref{sct:redshift} (PHZ, SPZ and CHM), and for arcs with
length-to-width ratio higher than both $d_0 = 7.5$ and $d_0 = 10$.
These quantities were then multiplied by the magnified cumulative
source number counts presented in Section \ref{sct:counts} (SM and
DB), to obtain the arc number counts as function of surface
brightness. The results, extrapolated to the whole sky, are shown in
Figure \ref{fig:arcCounts}. A detailed list with the predicted number
of submm and radio arcs for different sensitivities is presented in
Tables \ref{arcs_submm_all} and \ref{arcs_radio_all}, respectively.

Note that the arc number counts given by the source redshift
distributions SPZ and PHZ are almost indistinguishable on the scale of
Figure \ref{fig:arcCounts}, irrespective of the length-to-width
threshold $d_0$ adopted. As expected, SPZ produces more large arcs
than PHZ because it peaks at higher redshift, but only by a factor of
$\sim 7\%$. The redshift distribution CHM, on the other hand, produces
more large arcs than the other two by a factor of $\sim 2$. Since, as
mentioned in Section \ref{sct:redshift}, the distributions CHM and PHZ
can be considered as upper and lower limits to the true redshift
distribution respectively, we can conclude that the uncertainty
introduced by the redshift distribution in the predicted number of
arcs is less than a factor of two.

In terms of the length-to-width threshold, the number of arcs
predicted for $d_0 = 7.5$ is larger than for
$d_0 = 10$ (as was also expected). However, the ratio
between the number of arcs with $d \ge d_0 = 7.5$ and $d \ge d_0 = 10$
is not exactly equal to the ratio in the respective optical depths, since the
magnification distributions for the two kinds of arcs are also
different (see the discussion in \citealt{FE08.1}).

Finally, when it comes to comparing the results from the two adopted
source number counts (DB and SM), we see that the difference in the
predicted number of arcs is negligible for submm surface brightness
limits greater than $5$~mJy arcsec$^{-2}$. However, at
$B_{850~\mu\mathrm{m}}=0.5$~mJy arcsec$^{-2}$, the function DB
predicts $2$ times more arcs than SM, and the difference becomes a
factor of $5$ for $0.1$~mJy arcsec$^{-2}$. In the radio domain, the
difference between DB and SM is negligible for $B_{1.4 ~\mathrm{GHz}}
= 50~\mu$Jy arcsec$^{-2}$, a factor $\sim 2$ for $10~\mu$Jy
arcsec$^{-2}$, a factor $\sim 3$ for $5~\mu$Jy arcsec$^{-2}$ and a
factor $\sim 8$ for $1~\mu$Jy arcsec$^{-2}$. Therefore, the
uncertainty in the predicted number of arcs is clearly dominated by
the uncertainty of the source number counts at the faint surface
brightness end.

Considering an all-sky submm survey with enough resolution to resolve
individual arcs with $d \ge d_{0}=7.5$ ($\sim 0.2\arcsec$), and a
sensitivity of $B_{850~\mu\mathrm{m}} = 1$ mJy arcsec$^{-2}$, our
calculations predict between $\sim 500$ (PHZ SM) and $\sim 600$ (SPZ
DB) arcs with a signal-to-noise ratio (SNR) larger than
$5$. In the case of $d \ge d_{0} = 10$, the expected number of arcs
would be between $200$ and $250$. If the sensitivity is reduced to
$0.1$ mJy arcsec$^{-2}$, these predictions can vary between $\sim
3400$ and $\sim 8300$ for $d_{0}=7.5$, and between $1400$ and $3600$
for $d_{0}=10$.

In a similar way, an all sky radio survey with a sensitivity of
$B_{1.4 ~\mathrm{GHz}} = 0.1$~mJy arcsec$^{-2}$ would detect between 8
and 50 arcs for $d_{0}=7.5$ (between none and $25$ for $d_{0}=10$) at
SNR $\ge 5$, with the main uncertainty given by the
$S_{850~\mu\mathrm{m}}/S_{1.4 \mathrm{GHz}}$ ratio used to obtain the
radio number counts by scaling the observed submm counts. If the
limiting radio surface brightness is reduced to $\sim 10~\mu$Jy
arcsec$^{-2}$, the predicted number of arcs could be increased to
$\sim 500 - 1300$ for $d_{0}=7.5$ ($200 - 600$ for $d_{0}=10$). In
order to have excellent statistics with a few thousand giant arcs
with SNR $\ge 5$, it would be necessary to go as deep as
$S_{1.4 \mathrm{GHz}} = 1~\mu$Jy arcsec$^{-2}$. The largest
number of arcs is given by the CHM source redshift distribution, which
is about a factor of 2 larger than the number of arcs predicted by SPZ
and PHZ.

It is plausible that future radio and submillimeter surveys of galaxy
clusters would focus on the most massive objects, since the center of
attention of many multiwavelength studies is on X-ray bright
clusters. To roughly evaluate the effect of this kind of selection, we
re-computed the optical depths by including only those clusters in our
synthetic population with mass $M \ge 5 \times 10^{14} M_\odot
h^{-1}$. The resulting arc number counts are presented in Figure
\ref{fig:arcCounts_mass}, where we used the same scale and line types
as in Figure \ref{fig:arcCounts} to ease comparison. The corresponding
numbers of submm and radio arcs predicted for different sensitivities
are listed in Tables \ref{arcs_submm_massive} and
\ref{arcs_radio_massive}, respectively. The most interesting feature
about these new plots is that the relative difference in the arc
number counts produced by the CHM distribution on the one hand, and PHZ
and SPZ on the other, is significantly reduced. This is due to the
fact that massive clusters are found mainly at low redshift, hence
lower-mass higher-redshift lenses, that are accessible only to the CHM
distribution, become unimportant.

The order-of-magnitude reduction in the abundance of giant arcs when
focusing only on clusters with mass $M \ge 5 \times 10^{14} M_\odot
h^{-1}$ is in agreement with the fact that the bulk of the lensing
signal actually comes from low-mass clusters, since the optical depth
is obviously dominated by the lowest mass objects that are capable of
producing a non-vanishing cross sections (see Eq. \ref{eqn:tau}). Note
also that, unlike the previous case, PHZ produces slightly more arcs
than SPZ. This is due to the fact that we are including in the
calculations only low-$z$ clusters, and PHZ actually has more sources
with, e.g., $z_\mathrm{s} > 1$ than SPZ (see Figure \ref{fig:zd}).

Considering again an all-sky submm survey with enough resolution to
resolve individual arcs with $d_{0} = 7.5$ and sensitivity of
$B_{850~\mu\mathrm{m}} = 1$ mJy arcsec$^{-2}$, we predict $\sim 20$
arcs \textbf{with $\rm SNR \ge 5$} (8 if $d_{0}=10$). This number can
increase to $100 - 250$ $(40 - 90)$ if $B_{850~\mu\mathrm{m}} \ge 0.1$
mJy arcsec$^{-2}$. In the case of a radio survey, it would be
necessary to push the limiting flux density down to $\sim 1~\mu$Jy
arcsec$^{-2}$ to detect few hundred arcs (between $40 - 200$ if
$d_{0}=10$).

\section{Comparison with previous work}\label{sct:relations}

\begin{figure*}[t]
  \includegraphics[width=0.5\hsize]{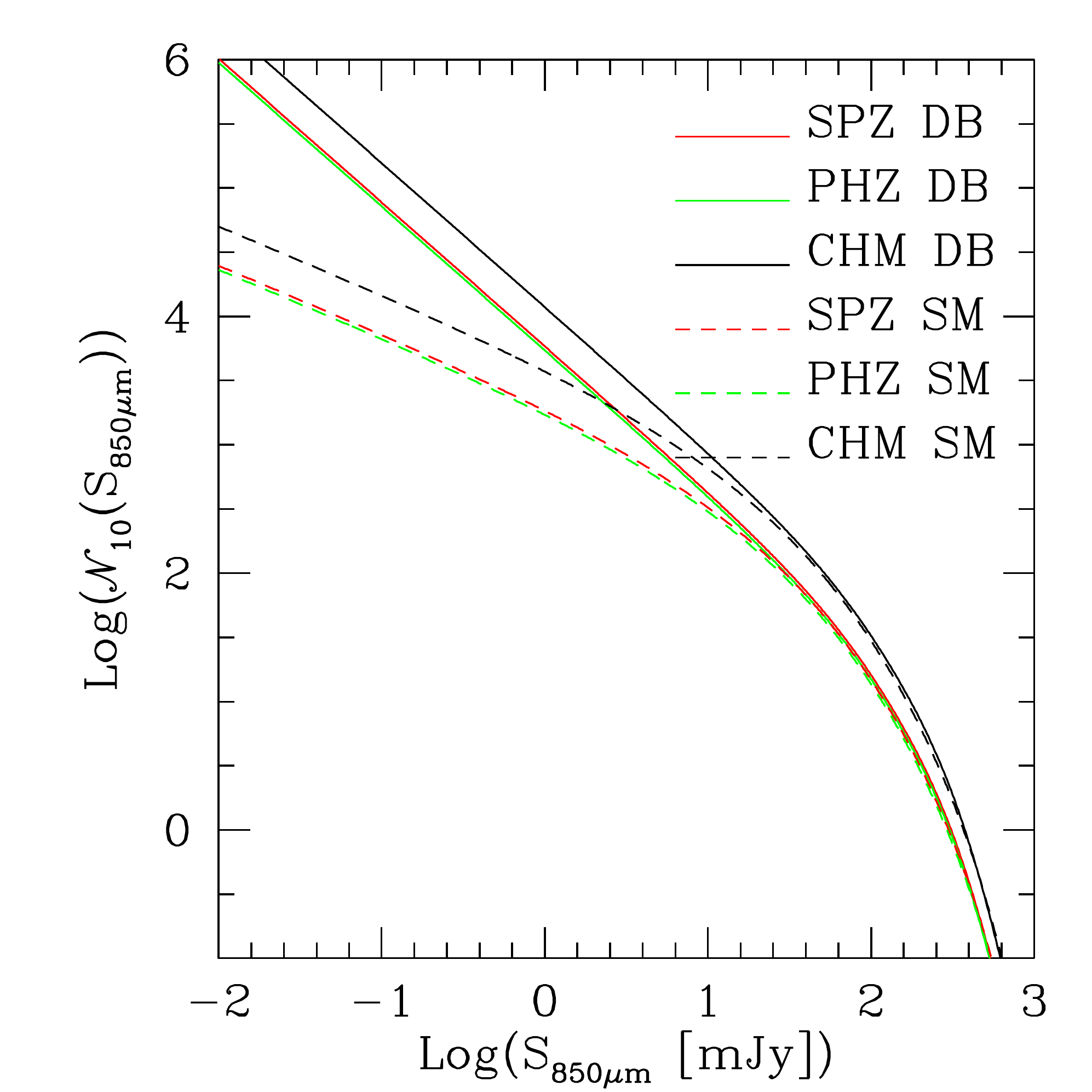}\hfill
  \includegraphics[width=0.5\hsize]{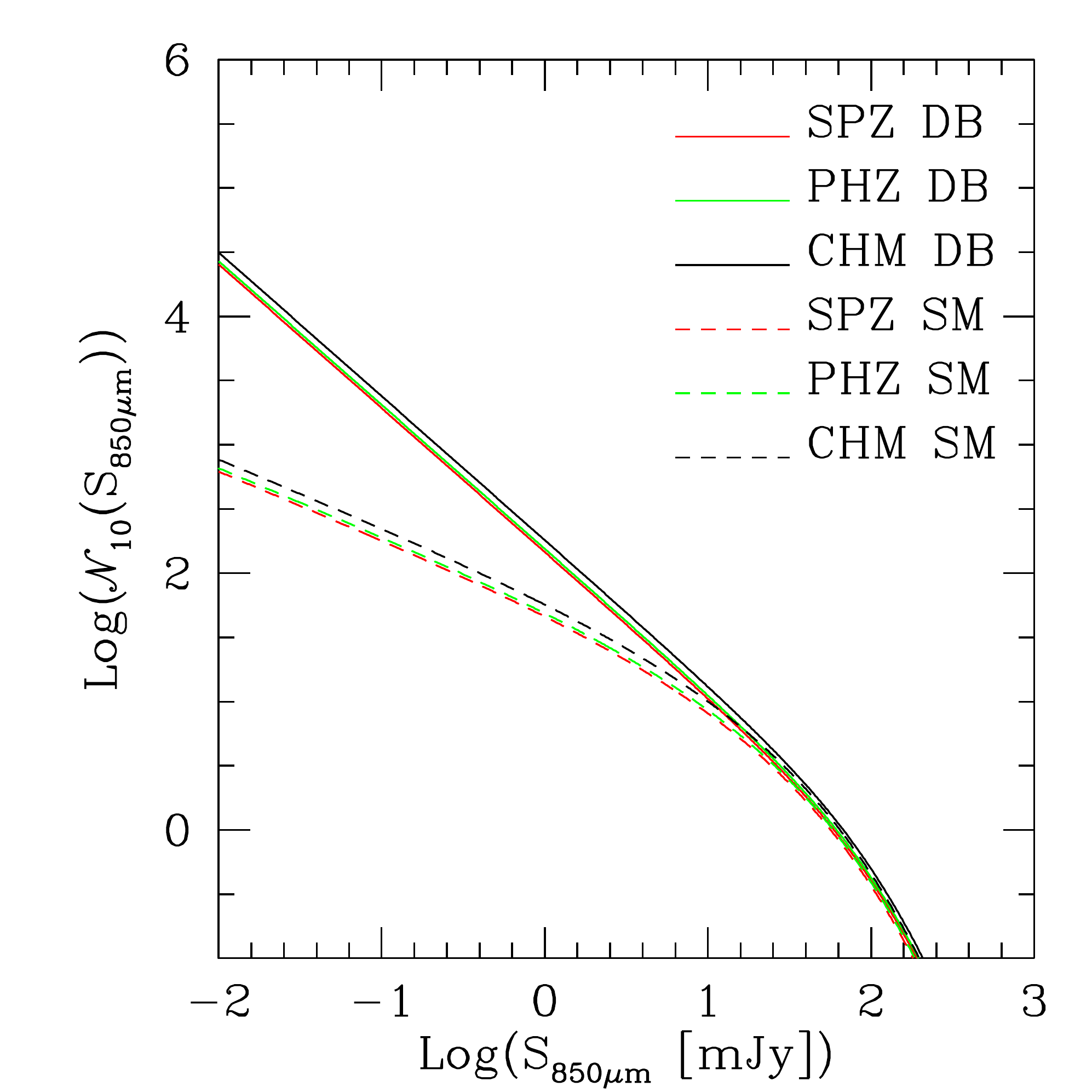}\hfill
  \includegraphics[width=0.5\hsize]{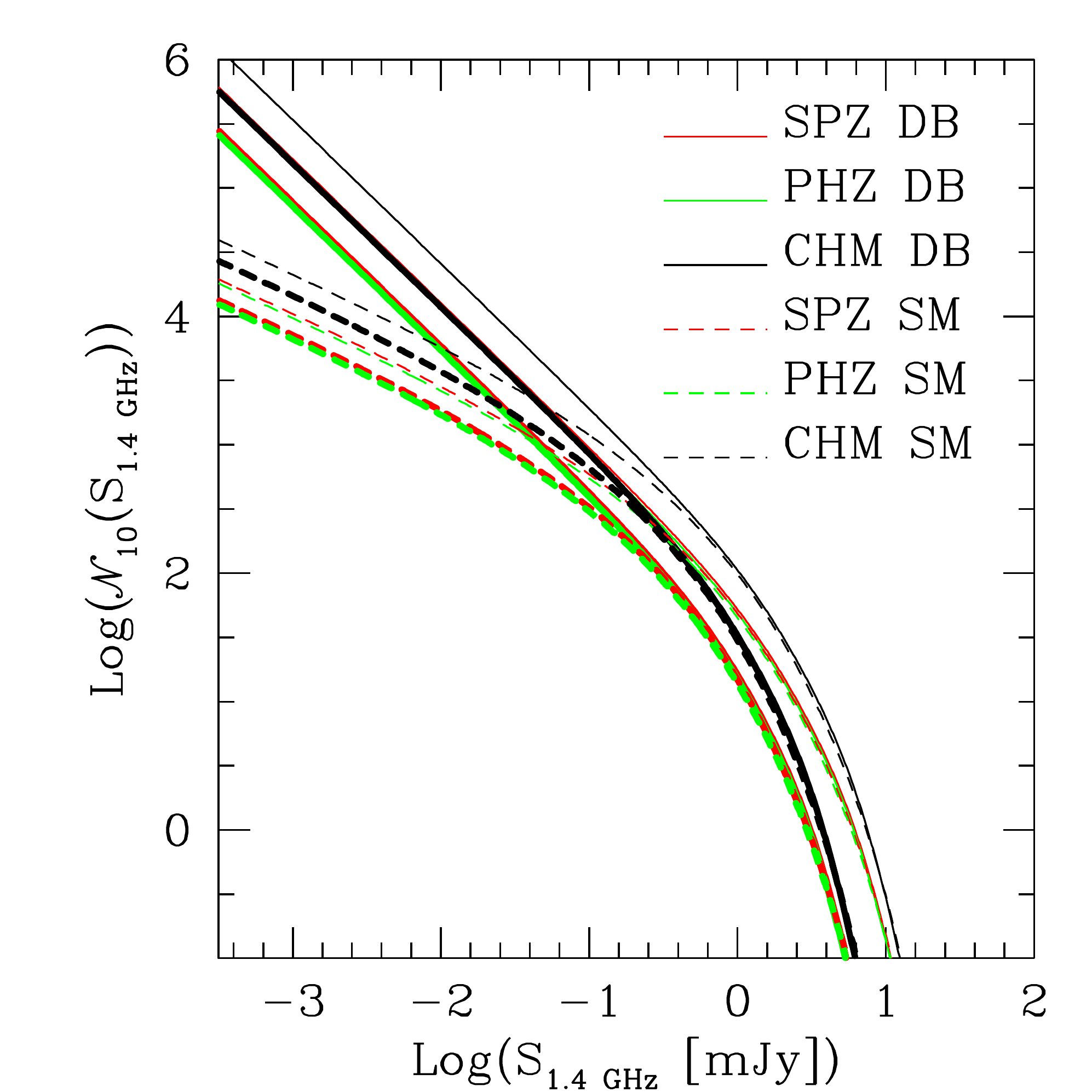}\hfill
  \includegraphics[width=0.5\hsize]{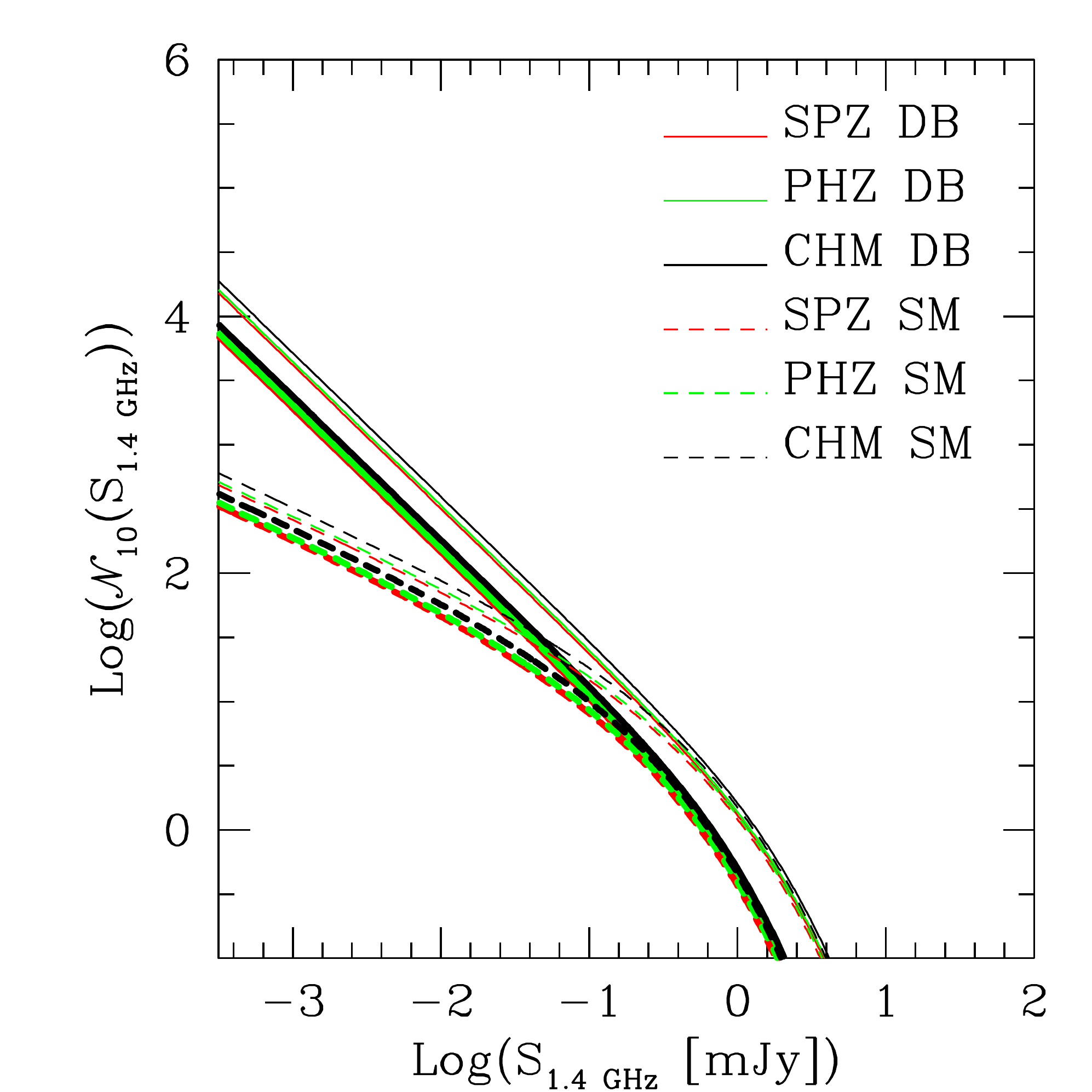}
\caption{The number of arcs with $d \ge d_0 = 10$ predicted to be
observed in the whole sky as a function of the submm flux density at
$850~\mu$m (top two panels) and radio flux density at 1.4~GHz (bottom
two). The left panels refer to all clusters in the sample, while the
right ones shows results when only clusters with mass $M \ge 10^{14}
M_\odot h^{-1}$ are considered. Different line styles and colors refer
to different number count models and source redshift distributions. In
the two bottom panels, the thin lines assume $S_{1.4 \mathrm{GHz}} =
S_{850~\mu \mathrm{m}}/50$, while the thick lines refer to $S_{1.4
\mathrm{GHz}} = S_{850~\mu \mathrm{m}}/100$.}
\label{fig:arcCounts_comparison}
\end{figure*}

The probability of strong gravitational lensing due to background
sources at radio and submm wavelengths has been a poorly studied issue
in the past years. In addition, the few works available in the
literature usually involve cosmological models, deflector mass ranges,
and modeling approaches for the source and lens populations that are
different from the ones used in the present work, making the
comparison between them difficult and often not possible. With this
note of caution, we tried however to make some of these tentative comparisons in
the following. This required us to repeat the calculations in the last
section using the number counts as a function of flux density instead
of surface brightness. The results are shown in Figure
\ref{fig:arcCounts_comparison} only for arcs with $d \ge d_0 = 10$.

\subsection{Submm wavelengths}

Given the poor resolution of current submm instruments, observational
studies of submm arcs have not been possible so far, with only one
submm arc candidate reported until now. This arc is supposed to be the
brightest region of the extended submm source SMM J$04542-0301$,
located in the core of the cluster MS$0451.6-0305$
(\citealt{BO04.1,BE07.1}; Berciano Alba et al. 2009, in
press). However, higher resolution submm observations are required in
order to confirm this hypothesis, and meaningful estimates of the
length and width of this source, unfortunately, cannot yet be done. On
the theoretical side, gravitational lensing of SMGs due to galaxy
clusters was first studied by \cite{BL97.1}, using a circularly
symmetric model of the cluster A2218 and the submm counts predicted by
different galaxy evolution models. More recent studies \citep{CO99.2,
PA09.1} also have been focused on predicting the number of submm
lensed sources, but predictions for the abundance of submm arcs have
never been attempted before.

For instance, in the work of \cite{PA09.1}, the authors employ the
strong lensing analysis of the \emph{Millennium Simulation} performed
by \cite{HI07.1} (see also \citealt{HI08.1}) in order to compute (i)
the average magnification of SMGs as a function of flux density, and
(ii) the contribution to the differential number counts given by
sources with different redshifts and magnifications. Hence, their
results cannot be compared with ours in a straightforward way.

The only work with which we could try a tentative comparison is the
one by \cite{CO99.2} (CO99 hereafter), where the author provides
number counts of (among others) gravitationally lensed submm sources
as a function of their magnification. The clusters were modeled as
Singular Isothermal Sphere (SIS henceforth) density profiles, which
means that the image magnification equals its length-to-width ratio,
as long as sources are circular and point-like. Since the sources that
we are using are not circular, nor point-like, the following
comparison should be taken with caution. The background submm
sources were described by means of the redshift and number
distributions observed in the Hubble Deep Field (HDF). Using a
$\Lambda$CDM cosmology, and considering cluster lenses with $M \ge 8.8
\times 10^{14} M_\odot h^{-1}$, CO99 predicted $\sim 500$ submm
sources in the whole sky with lensing magnification larger than $10$
and $S_{850~\mu m} \ge 2$ mJy.

On the other hand, our results indicate that, if no mass selection is
applied and for $S_{850~\mu m} \ge 2$ mJy, we should find a few
thousand arcs with $d \ge d_0 = 10$ in the whole sky. Restricting the
cluster mass range to $M \ge 5 \times 10^{14} M_\odot h^{-1}$ reduces
the number of arcs to $\sim 100$ at most. Matching the mass range of
CO99 would only reduce the number of predicted arcs even further,
hence being discrepant with CO99 predictions. Assuming that the source
population we are considering is the same, we ascribe at least part of
this disagreement to the fact that CO99 considers a very high
normalization of the power spectrum ($\sigma_8 \simeq 1.2$), which was
based on old studies of the cluster temperature function
\citep{VI96.1}. As shown in \cite{FE08.1}, this results in greater
strong lensing optical depth, and therefore to an over-prediction of
the number of lensed images.

\subsection{Radio wavelengths}

The first observational search for radio arcs in galaxy clusters dates
back to \cite{BA95.1}, where the authors considered a cluster sample
of 46 objects with $z \lesssim 0.3$ observed, among others, at a
wavelength of $20$ cm with a sensitivity of $\sim 1$ mJy. They claimed
a systematic tangential alignment of radio images with respect to the
cluster centers, and concluded that these are arclets or even giant
arcs produced by background flat-spectrum radio sources. While their
resolution was too poor to reliably measure arc
morphological properties, these findings were subsequently questioned
by \cite{AN98.1}, who performed a similar analysis on the
\cite{AB89.1} cluster sample (ACO henceforth) with sources taken from
the FIRST (Faint Images of the Radio Sky at Twenty-centimeters)
catalogue \citep{WH97.1}, finding no evidence for a preferential
alignment of radio images in the core of massive galaxy clusters. They
also report a rough estimate of the abundance of strong lensing events
that should be seen in ACO clusters, finding that for a statistically
significant detection, the limiting flux density should be lowered to
$\lesssim 0.1$ mJy, in agreement with previous, more detailed
estimates \citep{WU93.1}. More recently, \cite{PH01.1}
searched for strong lensing events with large angular separations in
the FIRST catalog, but none of their candidates turned out to be a
real gravitational lens.

About $40\%$ of the clusters included in the sample of \cite{BA95.1}
have velocity dispersion $\sim 1000$ km s$^{-1}$. Assuming that cluster
galaxies have the same velocity dispersion of the dark-matter
particles (\citealt{GA04.1,BI06.1,FA06.1}, see however
\citealt{CO00.1}), and adopting the simulation-calibrated scaling
relation of \cite{EV08.1}, this corresponds to a mass of at least
$\sim 8 \times 10^{14} M_\odot h^{-1}$. The rest of their clusters
should have a mass $\gtrsim 4 \times 10^{14} M_\odot h^{-1}$. Assuming
a limiting flux of $\sim 1$ mJy at $1.4$~GHz, our predictions in
Figure \ref{fig:arcCounts_comparison} give $\lesssim 100$ giant radio
arcs in the whole sky when the entire cluster population is
considered, and only $\lesssim 1$ inside clusters with mass $M \ge 5
\times 10^{14} M_\odot h^{-1}$. Therefore, we find it quite
unlikely that the detection claimed by \cite{BA95.1} is due to the
radio emission from SMGs.

The previous considerations suggest the need to go to flux densities
fainter than $1$ mJy in order to detect large radio arcs in galaxy
clusters. This kind of implication is also supported by the results of
\cite{CO98.1}, which showed that the radio emission corresponding to
large optical arcs in three out of the four cases he studied is $\lesssim 0.5$
mJy. In the only optical arc with a secure radio counterpart in that
work, namely arc $A0$ in cluster $A370$, the radio counterpart does not seem to be
arc-shaped, which may be due to the radio emitting region being offset
with respect with the optical emission or to resolution issues.

On the theoretical side, the statistics of radio arcs was
first investigated by \cite{WU93.1} (WU93 hereafter). Their
predictions were made using the evolutionary model of \cite{DU90.1} to
describe the radio luminosity function (dominated by starforming
galaxies and AGNs preferentially located at $z \lesssim 1$) at $2.7$~GHz
assuming a cosmological model with $\Omega_{\mathrm{m},0} = 1$. Conversely,
we focus on the radio counterparts of SMGs (preferentially
located at $z \sim 2$) at $1.4$~GHz in a more modern $\Lambda$CDM
cosmology. As pointed out in \cite{BA98.1}, the optical depth for optical
arcs produced by sources at $z_\mathrm{s} \sim 1$ grows by about one
order of magnitude in going from an Einstein-de Sitter universe to a
$\Lambda$CDM one. Therefore, we can use this prescription for comparison between
both predictions. Note that, since the SED of SMGs is rather flat at
radio wavelengths (spectral index $\sim 0.7$, \citealt{CO92.1}), we
expect only a small change in the flux density between $2.7$~GHz and
$1.4$~GHz.

In their calculations, WU93 only considered clusters with $z \lesssim
0.6$ and a galaxy velocity dispersion $> 800$ km s$^{-1}$, which
corresponds to a mass $\gtrsim 3.5 \times 10^{14} M_\odot
h^{-1}$. After applying the cosmology correction mentioned before,
their predicted number of giant radio arcs for the whole sky is (i)
$\sim 1000$ if $S_{2.7~\mathrm{GHz}} \gtrsim 10~\mu$Jy, (ii) $\lesssim
200$ if $S_{2.7~\mathrm{GHz}} > 0.1$~mJy and (iii) very small if
$S_{2.7~\mathrm{GHz}} > $ a few mJy (such that none should be detected
in surveys in the literature at that time). Considering only clusters
with mass $\ge 5 \times 10^{14} M_\odot h^{-1}$ (which have a redshift
range comparable to the one used in WU93), our predicted number of
radio arcs with $d \ge d_0 = 10$ produced by SMGs in the whole sky is
(i) $\sim 100 - 1000$ if $S_{1.4~\mathrm{GHz}} > 10~\mu$Jy, (ii) few
tens if $S_{1.4~\mathrm{GHz}} > 0.1$~mJy and (iii) very small if
$S_{1.4~\mathrm{GHz}} > $ a few mJy.

Our results may thus seem compatible with those of WU99, considering
that the number density of SMGs is certainly smaller than the number
density of the entire radio source population. This inference is
however not conclusive, since WU99 are using sources at rather
low-redshift. We cannot say whether considering their same redshift
distribution and number counts would lead to a discrepant result.

In CO99 there is also a study about strongly magnified radio sources,
analogous to the submm sources. It is found that $\sim 20$ sources
with amplification larger than $10$ should be found with
$S_{1.4~\mathrm{GHz}}\ge 10~\mu$Jy. For the same parameters, we find
in our high-mass cluster study a number ranging from $\sim 100$ up to
$\sim 1000$, which is much larger than the findings of CO99. This
discrepancy should be even more enhanced if the two calculations are
reduced to the same $\sigma_8$. The origin of this discrepancy is not
clear, although it might be related to the higher mass threshold that
they adopt, and the fact that the redshift distribution considered by
CO99 peaks at $z \sim 1$ instead of at $z \sim 2$. It should also be
noted that CO99 attribute their finding many fewer radio arcs than
WU93 to the different number count evolution adopted.

Our calculations of the
abundance of radio and submm arcs are more accurate than the works
discussed above for different reasons, mainly the different modeling
of the cluster population. To start with, in the literature, investigators
often consider all lenses as isolated, spherically symmetric density
distributions. \cite{WU93.1} investigate the effect of elliptical mass
distributions, but only on the magnification pattern and not on the
efficiency for the production of large arcs. On the other hand, we
included the effect of asymmetries, substructures and cluster mergers,
that all have been found to be important to augment arc
statistics. Next, we used an NFW density profile to model individual
lenses, which is a good representation of average dark-matter dominated
objects like galaxy clusters, while other works have often considered
SIS or SIS-like profiles, which are more suitable for galaxy
lensing. While for a SIS lens model the image magnification equals the
length-to-width ratio, it is known to produce fewer gravitational arcs
with respect to the more realistic NFW profile \citep{ME03.1}.

\section{Summary and conclusions}\label{sct:conclusions}

The advent of the high resolution submm facilities ALMA and CCAT, and
the major technological development that radio interferometry is
currently undergoing (e.g., \emph{e}-MERLIN, EVLA and SKA) will make possible the
study of radio and submm giant arcs produced by clusters of
galaxies. In particular, the study of giant arcs produced by submm
galaxies (SMGs) seems particularly promising for at least two reasons.

\begin{itemize}
\item[$\bullet$] It provides the opportunity to detect and spatially resolve the
morphologies and internal dynamics of this population of dust obscured
high-redshift star-forming galaxies, which is very difficult to study
in the optical.

\item[$\bullet$] It can provide information about the formation and evolution of
the high redshift cluster population, by means of arc statistics studies.
\end{itemize}

To assess the prospects for these kind of studies, we provided
theoretical predictions on the abundance of gravitational arcs
produced by the SMG population at radio ($1.4$~GHz) and submm ($850~
\mu$m) wavelengths, greatly improving the accuracy of the results with
respect to the first few studies carried out a decade (or more)
ago. The advantage of radio observations is that the angular
resolution and sensitivity provided by interferometers like
\emph{e}-MERLIN and EVLA are already (or will very soon be) at the
level required for these kind of studies. However, these frequencies
do not benefit from the same favorable K-correction as mm/submm
wavelengths do, which will make the latter a more interesting
tool for studying high-redshift clusters as soon as the resolution of
sub-mm observations is good enough. The calculation of the
number of arcs produced by a background source population requires
four main ingredients: (i) the source shape and size (ii), the source
redshift distribution, (iii) the cumulative source number counts, and
(iv) a model of the cluster population.

The model of the cluster population used in this work was based on the
extended \cite{PR74.1} formalism and made use of an NFW dentity
profile to describe each cluster lens. It also included the effect of
asymmetries, substructures and cluster mergers, which have been found
to play an important role in arc lensing statistics.

Based on current radio/CO observations and the FIR/radio correlation,
we have characterized the typical size of the radio and submm emitting
regions of SMGs with an effective radius $R_\mathrm{e}=0.25\arcsec$,
and an axis ratio that varies within the interval $b/a \in
[0.3,1]$. Resolving all the arcs produced by this kind of sources will
require $\sim 0.2\arcsec$ resolution.

Since the most accurate redshift distribution of SMGs available
\citep{CH05.1} is based on observations of the radio detected members
(which is biased against $z \ge 3$ sources), we used three different
functions to quantify the effect of a high redshift tail on the
predicted number of arcs. The results indicate that this effect is
less than a factor two if we consider all simulated clusters during
the calculations, and negligible if we only consider massive
clusters.

The submm source number counts used in this work correspond to the
joint fit of the (bright) SHADES survey counts and the (faint) Leiden
SCUBA Lens Survey counts presented in \cite{KN08.2}. To account for
the uncertainty in the low flux end, predictions were made for the
best fit to the data, and the shallowest fit consistent with the
data. Note that, although only $\sim 60\%$ of the observed SMGs have
being detected in radio, the next generation of radio interferometers
will be able to detect the radio counterparts of all SMGs with
$S_{850~\mu\mathrm{m}} \ge 5$~mJy. Therefore, the radio number counts
of SMGs have been derived by scaling the submm counts using
representative upper and lower limits of the
$S_{850~\mu\mathrm{m}}/S_{1.4~\mathrm{GHz}}$ ratio for SMGs at $z \sim
2$ taken from \cite{CH05.1}.

Our calculations show that a submm all-sky survey with a sensitivity
of $1$~mJy arcsec$^{-2}$ will detect hundreds of arcs with a $5\sigma$
significance. In the radio, this number can be achieved with a
sensitivity of $10-20~\mu$Jy arcsec$^{-2}$. Obtaining a statistically
significant sample of thousands of arcs would require sensitivities of
$0.1$~mJy arcsec$^{-2}$ in the submm and $1~\mu$Jy arcsec$^{-2}$ in the
radio. However, if only massive clusters ($M \ge 5 \times 10^{14}
M_\odot h^{-1}$) are considered in the calculations, the predicted
number of arcs is reduced by about an order of magnitude. 

Besides the many uncertainties involved in the theoretical predictions
presented here, the main challenge in designing a future survey for
radio/submm arc statistics studies will be in finding the best
compromise between survey area, depth and resolution, three issues
that affect the arc detectability in different manners. We believe
that this work provides a significant step forward in this direction.

\begin{table*}
\centering
\caption{Predicted number of submm ($850~\mu$m) arcs produced by SMGs
for an all-sky survey, using all the synthetic cluster population to
compute the optical depths (see Figure \ref{fig:arcCounts}).}
\label{arcs_submm_all}
\begin{tabular}{c|cc|cc|cc|cc}
\hline\hline
\noalign{\smallskip}
$B_{850~ \mu\mathrm{m}}^{~(1)}$    & \multicolumn{2}{c|}{$\rm 5~mJy ~arcsec^{-2}$} & \multicolumn{2}{c|}{$\rm 1~mJy ~arcsec^{-2}$} & \multicolumn{2}{c|}{$\rm 0.5~mJy ~arcsec^{-2}$} & \multicolumn{2}{c}{$\rm 0.1~mJy ~arcsec^{-2}$} \\
\noalign{\smallskip}
$d_{0}^{~(2)}$       &  $7.5$ & $10$        & $7.5$ & $10$  & $7.5$ & $10$  & $7.5$ & $10$ \\
\noalign{\smallskip}
\hline 
\noalign{\smallskip}
SPZ DB$^{~(3)}$ &575  &250  &3792  &1650  &8275   &3600  &50767  &22075  \\
PHZ DB$^{~(3)}$ &550  &233  &3575  &1533  &7800   &3342  &47892  &20517  \\
CHM DB$^{~(3)}$ &1142 &517  &7425  &3325  &16200  &7267  &99458  &44592  \\
\noalign{\smallskip}
\hline
\noalign{\smallskip}
SPZ SM$^{~(3)}$ &542   &233  & 2208 &958   &3567  &1550  &9658  &4200  \\
PHZ SM$^{~(3)}$ &508   &217  & 2083 &892   &3367  &1442  &9117  &3900  \\
CHM SM$^{~(3)}$ &1058  &475  & 4317 &1933  &6992  &3133  &1058  &475  \\
\noalign{\smallskip}
\hline
\noalign{\smallskip}
\multicolumn{9}{l}{$^{(1)}$ surface brightness limit used to determine the number of arcs.}\\
\multicolumn{9}{l}{$^{(2)}$ arc length-to-width ratio thresholds.}\\
\multicolumn{9}{l}{$^{(3)}$ redshift distribution and number count function, as introduced in Figures \ref{fig:zd} and \ref{fig:cc}.}\\
\end{tabular}
\end{table*}

\begin{table*}
\centering
\caption{Predicted number of submm ($850~\mu$m) arcs produced by SMGs
for an all-sky survey, using only clusters with $M \ge 5 \times
10^{14} ~M_\odot h^{-1}$ to compute the optical depths (see Figure
\ref{fig:arcCounts_mass}).}
\label{arcs_submm_massive}
\begin{tabular}{c|cc|cc|cc|cc}
\hline\hline
\noalign{\smallskip}
$B_{850~\mu\mathrm{m}}^{~(1)}$   & \multicolumn{2}{c|}{$\rm 5~mJy ~arcsec^{-2}$} & \multicolumn{2}{c|}{$\rm 1~mJy ~arcsec^{-2}$} & \multicolumn{2}{c|}{$\rm 0.5~mJy ~arcsec^{-2}$} & \multicolumn{2}{c}{$\rm 0.1~mJy ~arcsec^{-2}$} \\
\noalign{\smallskip}
$d_{0}^{~(2)}$       &  $7.5$ & $10$        & $7.5$ & $10$  & $7.5$ & $10$  & $7.5$ & $10$ \\
\noalign{\smallskip}
\hline 
\noalign{\smallskip}
SPZ DB$^{~(3)}$ &17 &8  &125  &42  &267  &92  &1633   &550  \\
PHZ DB$^{~(3)}$ &17 &8  &133  &42  &283  &92  &1733   &583  \\
CHM DB$^{~(3)}$ &25 &8  &150  &50  &325  &108  &2008  &683  \\
\noalign{\smallskip}
\hline
\noalign{\smallskip}
SPZ SM$^{~(3)}$ &17  &8 &75  &25  &117  &42  &308  &108  \\
PHZ SM$^{~(3)}$ &17  &8 &75  &25  &125  &42  &333  &108  \\
CHM SM$^{~(3)}$ &25  &8 &83  &33  &142  &50  &383   &133    \\
\noalign{\smallskip}
\hline
\noalign{\smallskip}
\multicolumn{9}{l}{$^{(1)}$ surface brightness limit used to determine the number of arcs.}\\
\multicolumn{9}{l}{$^{(2)}$ arc length-to-width ratio thresholds.}\\
\multicolumn{9}{l}{$^{(3)}$ redshift distribution and number count function, as introduced in Figures \ref{fig:zd} and \ref{fig:cc}.}\\

\end{tabular}
\end{table*}

\section*{Acknowledgments}

{\small We are grateful to M. Bartelmann, A. Blain and
L. V. E. Koopmans for reading the manuscript and for many useful
comments. We also would like to thank M. Swinbank, for providing us
the histograms presented in Figure \ref{fig:z_comparison}, their
Gaussian fits, and the predictions from his evolutionary model, and
K. K. Kundsen for providing us with the observational data presented
in Figure \ref{fig:counts_swinbank}. We also acknowledge stimulating
conversation with M. Bonamente, M. Brentjens, M. Joy, A. F. Loenen and
I. Prandoni. We wish to thank the anonymous referee for useful remarks
that allowed us to improve the presentation of our
work. C. F. acknowledges financial contributions from contracts
ASI-INAF I/023/05/0 and ASI-INAF I/088/06/0.}

{\small
\bibliographystyle{aa}
\bibliography{./master}
}

\onecolumn
\begin{landscape} 
\begin{table*}
\centering
\caption{Predicted number of radio (1.4~GHz) arcs produced by SMGs for
an all-sky survey, using all the synthetic cluster population to
compute the optical depths (see Figure \ref{fig:arcCounts}). Values
with and without brackets correspond to $S_{850~\mu\mathrm{m}} /
S_{1.4~\mathrm{GHz}}=100$ and $50$, respectively.}
\label{arcs_radio_all}
\begin{tabular}{@{\extracolsep{-1mm}}c|cc|cc|cc|cc|cc|cc}
\hline\hline
\noalign{\smallskip}
$B_{1.4~\mathrm{GHz}}^{~(1)}$   & \multicolumn{2}{c|}{$\rm 500~\mu Jy ~arcsec^{-2}$} & \multicolumn{2}{c|}{$\rm 100~\mu Jy ~arcsec^{-2}$} & \multicolumn{2}{c|}{$\rm 50~\mu Jy ~arcsec^{-2}$} & \multicolumn{2}{c|}{$\rm 10~\mu Jy ~arcsec^{-2}$} & \multicolumn{2}{c|}{$\rm 5~\mu Jy ~arcsec^{-2}$} & \multicolumn{2}{c}{$\rm 1~\mu Jy ~arcsec^{-2}$} \\
\noalign{\smallskip}
$d_{0}^{~(2)}$       &  $7.5$ & $10$        & $7.5$ & $10$  & $7.5$ & $10$  & $7.5$ & $10$  & $7.5$ & $10$  & $7.5$ & $10$ \\
\noalign{\smallskip}
\hline 
\noalign{\smallskip}
SPZ DB$^{~(3)}$ &50  (8)   &25 (0)  &583  (233)   &250 (100)  &1333 (575)  &575  (250)   &8275 (3792)   &3600 (1650)  &18283 (8275)  &7950  (3600)   &110017 (50767)  &47825 (22075)  \\
PHZ DB$^{~(3)}$ &50  (8)   &17 (0)  &550  (225)   &233 (92)   &1258 (550)  &542  (233)   &7800 (3575)   &3342 (1533)  &17242 (7800)  &7383  (3342)   &103783 (47892)  &44450 (20517)  \\
CHM DB$^{~(3)}$ &100 (17)  &42 (8)  &1142 (458)   &517 (208)  &2608 (1142) &1167 (517)  &16200 (7425)   &7267 (3325)  &35808 (16200) &16058 (7267)   &215517 (99458)  &96633 (44592)  \\
\noalign{\smallskip}
\hline
\noalign{\smallskip}
SPZ SM$^{~(3)}$ &42 (8)  &17 (0)  &542  (225)   &233 (100)  &1058 (542)  &458 (233)  &3567 (2208)  &1550 (958)  &5608  (3567)   &2433 (1567)  &14325 (9658)  &2394  (4200)  \\
PHZ SM$^{~(3)}$ &42 (8)  &17 (0)  &508  (208)   &217 (92)   &1000 (508)  &425 (217)  &3367 (2083)  &1442 (892)  &5292  (3367)   &2267 (1442)  &13508 (9117)  &5783  (3900)  \\
CHM SM$^{~(3)}$ &83 (8)  &42 (8)  &1058 (442)   &475 (200)  &2075 (1058) &933 (475)  &6992 (4317)  &3133 (1933) &10983 (6992)   &492  (3133)  &28058 (18925) &12583 (8483)  \\
\noalign{\smallskip}
\hline
\noalign{\smallskip}
\multicolumn{9}{l}{$^{(1)}$ surface brightness limit used to determine the number of arcs.}\\
\multicolumn{9}{l}{$^{(2)}$ arc length-to-width ratio thresholds.}\\
\multicolumn{9}{l}{$^{(3)}$ redshift distribution and number count function, as introduced in Figures \ref{fig:zd} and \ref{fig:cc}.}\\
\end{tabular}
\end{table*}

\begin{table*}
\centering
\caption{Predicted number of radio (1.4~GHz) arcs produced by SMGs for
an all-sky survey, using only clusters with $M \ge 5 \times 10^{14}
~M_\odot h^{-1}$ to compute the optical depths (see Figure
\ref{fig:arcCounts_mass}). Values with and without brackets correspond
to $S_{850~\mu\mathrm{m}} / S_{1.4~\mathrm{GHz}}=100$ and $50$,
respectively.}
\label{arcs_radio_massive}
\begin{tabular}{c|cc|cc|cc|cc|cc|cc}
\hline\hline
\noalign{\smallskip}
$B_{1.4~\mathrm{GHz}}^{~(1)}$   & \multicolumn{2}{c|}{$\rm 500~\mu Jy ~arcsec^{-2}$} & \multicolumn{2}{c|}{$\rm 100~\mu Jy ~arcsec^{-2}$} & \multicolumn{2}{c|}{$\rm 50~\mu Jy ~arcsec^{-2}$} & \multicolumn{2}{c|}{$\rm 10~\mu Jy ~arcsec^{-2}$} & \multicolumn{2}{c|}{$\rm 5~\mu Jy ~arcsec^{-2}$} & \multicolumn{2}{c}{$\rm 1~\mu Jy ~arcsec^{-2}$} \\
\noalign{\smallskip}
$d_{0}^{~(2)}$       &  $7.5$ & $10$        & $7.5$ & $10$  & $7.5$ & $10$  & $7.5$ & $10$  & $7.5$ & $10$  & $7.5$ & $10$ \\
\noalign{\smallskip}
\hline 
\noalign{\smallskip}
SPZ DB$^{~(3)}$ &0 (0)  &0 (0)  &17 (8)  &8 (0)  &42 (17)  &17 (8)  &267 (125)  &92  (42)   &592 (267)  &200 (92)  &3542 (1633)  &1192 (550)   \\
PHZ DB$^{~(3)}$ &0 (0)  &0 (0)  &17 (8)  &8 (0)  &42 (17)  &17 (8)  &283 (133)  &92  (42)   &625 (283)  &208 (92)  &3758 (1733)  &1267 (583)  \\
CHM DB$^{~(3)}$ &0 (0)  &0 (0)  &25 (8)  &8 (0)  &50 (25)  &17 (8)  &325 (150)  &108 (50)   &725 (325)  &242 (108) &4350 (2008)  &1475 (683)  \\
\noalign{\smallskip}
\hline
\noalign{\smallskip}
SPZ SM$^{~(3)}$ &0 (0)  &0 (0)  &17 (8)  &8 (0)  &33 (17)  &8  (8)  &117 (75)  &42 (25)  &183 (117)  &58 (42)  &458 (308)  &158 (108)  \\
PHZ SM$^{~(3)}$ &0 (0)  &0 (0)  &17 (8)  &8 (0)  &33 (17)  &8  (8)  &125 (75)  &42 (25)  &192 (125)  &67 (42)  &492 (333)  &167 (108)  \\
CHM SM$^{~(3)}$ &0 (0)  &0 (0)  &25 (8)  &8 (0)  &42 (25)  &17 (8)  &142 (83)  &50 (33)  &225 (142)  &75 (50)  &567 (383)  &192 (133)  \\
\noalign{\smallskip}
\hline
\noalign{\smallskip}
\multicolumn{9}{l}{$^{(1)}$ surface brightness limit used to determine the number of arcs.}\\
\multicolumn{9}{l}{$^{(2)}$ arc length-to-width ratio thresholds.}\\
\multicolumn{9}{l}{$^{(3)}$ redshift distribution and number count function, as introduced in Figures \ref{fig:zd} and \ref{fig:cc}.}\\

\end{tabular}
\end{table*}
\end{landscape}
\twocolumn

\end{document}